\documentclass[final,authoryear,1p,times,twocolumn]{elsarticle}

\usepackage{graphicx}
\usepackage{longtable}

\biboptions{authoryear}

\journal{Planetary and Space Science}

\begin{document}

\begin{frontmatter}



\title{EURONEAR - Recovery, Follow-up and Discovery of NEAs and MBAs using Large Field 1-2m Telescopes\tnoteref{label1}}
\tnotetext[label1]{Based on observations taken with the telescopes ESO/MPG 2.2m in La Silla (ESO Run number 080.C-2003), Swope 1m in Las Campanas (CNTAC 2008) and the INT 2.5m in La Palma (CAT DDT 2010). }


\author[inst1,inst2,inst3,inst4]{O. Vaduvescu\corref{cor1}}
\ead{ovidiuv@ing.iac.es}

\author[inst2,inst5]{M.~Birlan}
\author[inst6,inst7,inst8,inst9]{A.~Tudorica}
\author[inst10,inst11]{A.~Sonka}
\author[inst3]{F.~Pozo~N.}
\author[inst3]{A.~Barr~D.}
\author[inst12]{D.~J.~Asher}
\author[inst4,inst13]{J.~Licandro}
\author[inst14]{J.~L.~Ortiz}
\author[inst3]{E.~Unda-Sanzana}
\author[inst2,inst11,inst15]{M.~Popescu}
\author[inst2,inst5]{A.~Nedelcu}
\author[inst8,inst11]{D.~Dumitru}
\author[inst7,inst8,inst16]{R.~Toma}
\author[inst17]{I.~Comsa}
\author[inst8]{C.~Vancea}
\author[inst11]{D.~Vidican}
\author[inst11]{C.~Opriseanu}
\author[inst8]{T.~Badescu}
\author[inst8]{M.~Badea}
\author[inst11]{M.~Constantinescu}

\cortext[cor1]{Corresponding author}

\address[inst1]{\scriptsize {Isaac Newton Group of Telescopes (ING), Apartado de Correos 321, E-38700 Santa Cruz de la Palma, Canary Islands, Spain}}
\address[inst2]{\scriptsize {IMCCE, Observatoire de Paris, 77 Avenue Denfert-Rochereau, 75014 Paris Cedex, France}}
\address[inst3]{\scriptsize {Instituto de Astronom\'ia, Universidad Cat\'olica del Norte (IA/UCN), Avenida Angamos 0610, Antofagasta, Chile}}
\address[inst4]{\scriptsize {Instituto de Astrof\'isica de Canarias (IAC), C/V\'ia L\'actea s/n, 38205 La Laguna, Spain}}
\address[inst5]{\scriptsize {Astronomical Institute of the Romanian Academy, Cutitul de Argint 5, Bucharest 040557, Romania}}
\address[inst6]{\scriptsize {Bonn Cologne Graduate School of Physics and Astronomy, Germany}}
\address[inst7]{\scriptsize {Argelander-Institut f\"ur Astronomie, Universit\"at Bonn, Auf dem H\"ugel 71 D-53121 Bonn, Germany}}
\address[inst8]{\scriptsize {University of Bucharest, Department of Physics, CP Mg-11, Bucharest Magurele 76900, Romania}}
\address[inst9]{\scriptsize {Institute for Space Sciences, Bucharest - Magurele, Ro-077125 Romania}}
\address[inst10]{\scriptsize {Astronomical Observatory ``Admiral Vasile Urseanu'', B-dul Lascar Catargiu 21, Bucharest, Romania}}
\address[inst11]{\scriptsize {Bucharest Astroclub, B-dul Lascar Catargiu 21, sect 1, Bucharest, Romania}}
\address[inst12]{\scriptsize {Armagh Observatory, College Hill, Armagh BT61 9DG, UK}}
\address[inst13]{\scriptsize {Departamento de Astrof\'isica, Universidad de La Laguna, E-38205 La Laguna, Tenerife, Spain}}
\address[inst14]{\scriptsize {Instituto de Astrof\'isica de Andaluc\'ia (IAA), CSIC, Apt 3004, 18080 Granada, Spain}}
\address[inst15]{\scriptsize {Polytechnic University of Bucharest, Faculty of Applied Sciences, Department of Physics, Bucharest, Romania}}
\address[inst16]{\scriptsize {Romanian Society for Meteors and Astronomy (SARM), CP 14 OP 1, 130170, Targoviste, Romania}}
\address[inst17]{\scriptsize {Babes-Bolyai University, Faculty of Mathematics and Informatics, 400084 Cluj-Napoca, Romania}}

\begin{abstract}
\small
We report on the follow-up and recovery of 100 program NEAs, PHAs and VIs using the 
ESO/MPG 2.2m, Swope 1m and INT 2.5m telescopes equipped with large field cameras. 
The 127 fields observed during 11 nights covered 29 square degrees. Using these data, 
we present the incidental survey work which includes 558 known MBAs and 628 unknown 
moving objects mostly consistent with MBAs from which 58 objects became official 
discoveries. 
We planned the runs using six criteria and four servers which focus mostly on
faint and poorly observed objects in need of confirmation, follow-up and recovery.
We followed 62 faint NEAs within one month after discovery and we recovered 10 faint 
NEAs having big uncertainties at their second or later opposition. Using the INT we 
eliminated 4 PHA candidates and VIs. We observed in total 1,286 moving objects and we reported 
more than 10,000 positions. All data were reduced by the members of our network in a 
team effort, and reported promptly to the MPC. The positions of the program NEAs 
were published in 27 MPC and MPEC references and used to improve their orbits. 
The O--C residuals for known MBAs and program NEAs are smallest for the ESO/MPG and 
Swope and about four times larger for the INT whose field is more distorted. For the 
astrometric reduction, the UCAC-2 catalog is recommended instead of USNO-B1. 
The incidental survey allowed us to study statistics of the MBA and NEA populations 
observable today with 1--2m facilities. We calculate preliminary orbits for all unknown 
objects, classifying them as official discoveries, later identifications and unknown 
outstanding objects. The orbital elements $a$, $e$, $i$ calculated by FIND\_ORB software 
for the official discoveries and later identified objects are very similar with the 
published elements which take into account longer observational arcs; thus preliminary 
orbits were used in statistics for the whole unknown dataset. 
We present a basic model which can be used to distinguish between MBAs and 
potential NEAs in any sky survey. Based on three evaluation methods, most of our unknown 
objects are consistent with MBAs, while up to 16 unknown objects could represent NEO 
candidates and four represent our best NEO candidates. 
We assessed the observability of the unknown MBA and NEA populations using 1m and 2m 
surveys. Employing a 1m facility, one can observe today fewer unknown objects than known 
MBAs and very few new NEOs. Using a 2m facility, a slightly larger number of unknown 
than known asteroids could be detected in the main belt. Between 0.1 and
0.8 new NEO candidates per square degree could be discovered using a 2m telescope. 

\end{abstract}


\end{frontmatter}

% \linenumbers

\section{Introduction}
\label{intro}

Near Earth Asteroids (NEAs) are defined as minor planets with a perihelion distance $q\leq1.3$ AU and an aphelion 
distance $Q\geq0.983$ AU \citep{mor02}. Potentially Hazardous Asteroids (PHAs) are NEAs having a minimum 
orbital intersection distance $MOID\leq0.05$ AU and an absolute magnitude $H\leq22$ mag \citep{bm94}. Virtual 
Impactors (VIs) represent objects for which the future Earth impact probability is non-zero according to the actual 
orbital uncertainty \citep{mg09}. 

According to present data \citep[e.g.][]{bow11}, there are more than half a million orbits of known Main Belt 
Asteroids (MBAs) and about 7,600 catalogued NEAs of which about 1,200 are PHAs \citep{nas11} and ca 100 VIs \citep{neo11}. 
During the last two decades the total numbers of discovered NEAs and PHAs have continued to grow, mainly thanks to 
five dedicated surveys led by the United States (CSS, LINEAR, Spacewatch, LONEOS and NEAT) which have been using 
large field mostly 1m class telescopes. Together, they have discovered $86.8\%$ of the entire NEA population known 
today (MPC Jan 2011 database), while Europe accounts for less than $1\%$ (led by La Sagra and Crni Vrh mostly run by
amateurs). 
Some European initiatives were taken by national institutions or local collaborations (ASIAGO/ADAS in Italy 
and Germany, CINEOS in Italy, KLENOT in the Czech Republic, NEON in Finland) and a few programs to study physical 
properties of NEAs have been carried out by groups in Europe (led by P. Pravec in the Czech Republic, SINEO led 
by M. Lazzarin in Italy, another program led by J. Licandro in Spain, etc).
Thus although there is still no common European program and no dedicated
telescope to observe NEAs, the existence of this expertise, in addition to
extensive observational facilities, evidently provides an opportunity for
Europe to improve its number of discoveries. 

In spite of the larger facilities apparently available today, extremely few researchers have used 2m class or 
larger facilities to observe fainter NEAs. During two short runs at ESO La Silla, \citet{boa04} employed the 
ESO/MPG 2.2m as a search facility and the NTT 3.5m as a follow-up telescope to survey faint NEAs and MBAs beyond 
22nd magnitude. During three nights using the ESO/MPG facility, the authors incidentally observed about 700 
MBAs as faint as $R=21.5$ mag ($V\sim22$ mag) exposing between 60 and 150 sec in the $R$ band. Using only four hours in 
override service mode at the Yepun VLT 8.2m telescope, \citet{boa03} eliminated 4 very faint VIs ($22 < V < 25$ mag), 
shifting them to simply the NEA or PHA class. 

Although the known NEA and PHA populations have increased during the last few decades, the annual growth of the number 
of known NEAs and PHAs appears to be becoming constant during recent years \citep{earn11}, probing some size threshold 
due to the limiting magnitude $V\sim21$ mag reached by the present 1m class surveys. It is unclear whether new 2m surveys 
such as the Spacewatch 1.8m and especially the new Pan-STARRS PS1 1.8m (although not entirely dedicated to NEAs) will 
make a significant contribution to the completeness of NEAs through the small size objects. 

During the last four years, the European Near Earth Asteroids Research (EURONEAR) program has observed to date 
234 program NEAs, defined as NEAs specifically planned to be observed according to a few selection criteria to
be discussed below. Throughout our program, we used 10 mostly 1m class telescopes, in visiting mode, contributing with 
follow-up astrometry and recovery of some important NEAs, PHAs and VIs, allowing their orbits to be secured or
improved \citep{bir10}. Part of this work, three telescopes in Chile and the Canaries, represent our largest
facilities employed, namely the ESO/MPG 2.2m in La Silla, Swope 1m in Las Campanas, and the INT 2.5m in La Palma, all 
equipped with large field cameras. To take advantage of these facilities, incidentally to our main program NEA 
work, we have identified many known MBAs in the observed fields and have also discovered many new objects. 

In this paper we review our EURONEAR observations at ESO/MPG and Swope and we present our new observations 
at the INT. Besides presenting our main NEA follow-up program, we introduce and discuss our incidental survey
work using the observed fields, classifying all the observed sources as known MBAs, unknown MBAs and NEA candidates. 
By comparing the statistics obtained from our survey on the three 1--2m facilities, and also with those obtained 
from other authors using 4m and 8m telescopes, we assess the observability limits of the MBA and NEA populations
observable with 1m and 2m telescopes. In Section~\ref{obsred} of the paper we 
explain the basic principles driving the planning, observations and data reduction of our runs. In 
Section~\ref{results} we present the results, including program NEAs, known MBAs and other unknown objects. 
In Section~\ref{disc} we discuss the results, focusing on the known and unknown MBA and NEA populations,
comparing all three facilities and presenting some statistics. The conclusions are presented in Section~\ref{conclusions}.

\section{Observations and Data Reduction}
\label{obsred}

As part of the EURONEAR project, between 2008 and 2010 we obtained three runs and a total of 11 nights for 
proposals devoted to the recovery and follow-up of some important NEAs, PHAs and VIs using the ESO/MPG and 
Swope telescopes in Chile and the INT in La Palma. For each run, the targets were selected based on the 
daily updated known NEA population. 

\subsection{Planning the Runs}

To search and prioritize the objects, we have used four planning servers with which one can check on 
a daily or hourly basis the updated NEA database. To complement the existing surveys focused on discovery, 
we focused our EURONEAR runs on some important objects in need of orbital improvement, especially on
newly discovered faint asteroids which need to be secured against loss and also on other older objects 
having a short observed arc which needs to be improved at the second or a later opposition. We selected 
our targets taking into account the following six {\it criteria}: 

\begin{enumerate}
\item 
Object class: observe objects classified as VIs, PHAs or NEAs (in this order), to improve their 
orbits and confirm or change their classification; 
\item 
Time interval from discovery: secure and follow-up poorly observed objects, a few days or weeks 
from discovery; 
\item 
Number of oppositions: recover objects previously observed at very few oppositions (especially
one-opposition objects); 
\item 
Object brightness: recover and follow-up faint and very faint objects, accessible only to larger 
facilities (larger than 2m) in danger of being lost by current surveys; 
\item 
Positional uncertainty: recover and follow-up poorly observed objects having large positional 
uncertainty (up to one degree), less accessible to other smaller field facilities; 
\item 
New object confirmation: recover newly discovered objects, preferably a few to several hours after their 
discovery; 
\end{enumerate}

To implement these criteria, we have used the following four {\it planning servers}: 

\begin{enumerate}
\item 
EURONEAR Planner 1: queries the Spaceguard database for mostly newly discovered objects; 
\item 
EURONEAR Planner 2: queries the MPC Bright and Faint Recovery Opportunity databases for mostly older 
objects in need of being recovered at a new opposition; 
\item
MPC NEO Confirmation Page: includes 1-night objects in need of being confirmed by independent observers; 
\item 
NASA/JPL Close Approaches List: includes closest approaches of old and new objects visible from Earth; 
\end{enumerate}

Both EURONEAR planner web-services are accessible online \citep{eur11}, being written in PHP by our team 
and offered to the community for planning other NEA follow-up campaigns. The servers query the current 
Spaceguard or MPC databases (both updated on a daily basis) and return the prioritized observing lists 
given the observing place, facility and observing date. The planning is based on eight calculated 
observability factors, namely: the asteroid class (according to MPC), the apparent magnitude, proper 
motion, ephemeris uncertainty, 
altitude, star density in the field, and Moon illumination and distance, calculated with a time step 
(e.g., one hour) in a given time interval (e.g., one night). The result consists of a few tables listing 
at each step the recommended observable objects prioritized according to the object apparent magnitude, 
its altitude (or airmass), proper motion, sky plane error, or some proposed ``Observability'' factor 
calculated as the product of all the above individual observability factors. Two accurate ephemeris servers, 
namely NEODyS \citep{neo11} and IMCCE \citep{imc11}, are queried by the planning server automatically, 
returning sky coordinates, magnitudes and uncertainties according to the observed orbital arc. 
The response time for both servers is short, usually less than one minute for one night's information. 

\subsection{Observations}

We present here our observing runs at ESO/MPG in La Silla, Swope in Las Campanas and INT in La Palma. 

\subsubsection{ESO/MPG Observing Run}

During 3 nights from 10--13 Mar 2008 we used the ESO/MPG 2.2m telescope at ESO La Silla, Chile 
to observe 15 program NEAs, PHAs and VIs. At the Cassegrain $F/5.9$ focus of the telescope we 
used the Wide Field Imager (WFI) which consists of a $2\times4$ mosaic of CCDs $2K\times4K$ pixels each, 
covering a total field of view of $34^\prime \times 33^\prime$ with a pixel scale of 
$0.24~^{\prime\prime}$/pix. 

Due to the relatively high proper motion of NEAs at opposition (around $2-3~^{\prime\prime}/$min), 
the average seeing of about $1-1.5^{\prime\prime}$, and taking into account the large raw image 
size on disk (140 MB) combined with the relatively slow readout time compared with our fast planned 
cadence, we observed the entire run in binning mode $3\times3$ ($0.71^{\prime\prime}$ pixel size). 
We used an $R$ band filter for the entire run. This binning does not affect the quality of our 
astrometry (set by the goal of having astrometric errors less than $0.3^{\prime\prime}$, i.e., 
comparable with the star catalog reference), as can be observed from the statistics in Section~\ref{results}. 

To take advantage of the large MPG and WFI facility, we focused our run on two aspects, namely 
to follow-up some important NEAs, and to discover and recover many MBAs appearing in the observed 
fields. The weather was clear all three nights, with only two hours lost due to high humidity at 
the beginning of the first night. The sky was dark, with the Moon 3--6 days past new.

In Table A.1. of our past paper \citep{bir10} we listed the 15 observed NEAs during our ESO/MPG run, 
6 VIs, 4 PHAs and 5 other NEAs. Besides the 15 NEA program fields, during the following available nights we 
observed nine neighbouring fields, in order to secure some MBAs discovered in the previous nights. The
neighbouring fields were chosen assuming a proper motion of $0.7~^{\prime\prime}/$min for MBAs 
observed near opposition. During the second night only, we also surveyed 8 WFI fields (2.5 square 
degrees) in the ecliptic, about $50^{\circ}$ from opposition to avoid crowding from the Milky Way. 
Besides the program NEAs, we identified and measured all moving sources in all the observed fields, 
reporting all known and new objects visible up to $V \sim 22$ mag. During all three nights at ESO/MPG, 
we observed in total 42 WFI fields covering about $13$ square degrees. 

\subsubsection{Swope Observing Run}

During 5 nights on 18--19 Oct and 22--24 Oct 2008 we used the Swope 1m telescope in Las Campanas 
Observatory (LCO), Chile, to observe 50 program NEAs. At the Cassegrain $F/7$ focus of the 
telescope we used the $SITe\#3$ $2K\times3.6K$ pixel camera giving a field of $15.1^\prime 
\times 26.5^\prime$ with a pixel scale $0.43~^{\prime\prime}$/pix. We used an $R$ band filter and 
no binning for the entire run. The sky was gray (Moon up to 4 days from last quarter) and the 
weather was very good, with seeing around $1^{\prime\prime}$. 

In Table A.1. of our past paper \citep{bir10} we listed our 50 observed NEAs during the Swope 
run, 12 PHAs and 38 other NEAs. Besides the program NEAs, we identified and measured all moving sources
in the observed fields, reporting all known and new objects visible up to $R \leq 20.4$ mag. During 
five nights we observed in total 50 Swope fields covering some 6 square degrees. 

\subsubsection{INT Observing Runs}

During two discretionary nights (D-nights) on 12 Feb and 15 Apr 2009 (3+3 hours), one hour in 
13/14 Nov 2009 and 4 nights awarded by the Spanish Director's Discretionary Time (DDT) in 20--25 
Feb and 3 Mar 2010, we used the INT 2.5m telescope in Roque de Los Muchachos Observatory (ORM) in 
La Palma to observe 35 program NEAs, namely 1 VI, 13 PHAs and 21 other NEAs. At the prime focus 
of the INT we used the Wide Field Camera (WFC) which consists of 4 CCDs $2K\times4K$ pixels each, 
covering an L-shape $34^\prime \times 34^\prime$ with a pixel scale of $0.33~^{\prime\prime}$/pix. 
Both D-nights and 13/14 Nov 2009 were observed without binning, while the Feb-Mar 2010 run was 
observed with $2\times2$ binning ($0.66~^{\prime\prime}$/pix) to minimize the readout time 
and match the poor weather conditions. We used an $R$ filter for all runs. 

Most of the INT time was bright and gray, with only three hours dark time. The weather was 
good during the first 6 hours on the first two D-nights and one hour in 13/14 Nov 2009, but
very bad (wet and windy) during the whole DDT run (Feb-Mar 2010) when the Moon was gray and 
bright. In Table~\ref{table1} we list the INT program NEAs, including the object classification, 
proper motion, $3\sigma$ positional uncertainty (according to the MPC for the observing date) and 
the orbital arc length at the observing date. Besides the program NEAs, at the INT we identified 
and measured all moving sources in all observed fields, reporting all known and new objects 
visible up to $R\sim21.2$ mag. In total at the INT, we observed for about three clear nights, 
a total of 35 WFC fields covering some 10 square degrees.

\subsection{Data Reduction}

For all runs, we processed the data within 2--3 days from observations using an IRAF pipeline 
(for the INT and Swope data) and IDL (for ESO/MPG data), taking into account the usual subtraction 
of the appropriate bias and sky flat field. For the WFI and WFC mosaic cameras we sliced CCD images 
and treated them independently, calculating CCD centres based on the pointing of the telescope 
(included in the image headers) and the geometry of the two cameras. All images were processed 
on-site, then archived and eventually transferred via FTP to the remote available data reducers,
allocated on demand. 

\begin{figure*}
  \centerline{
    \mbox{\includegraphics[width=7.5cm]{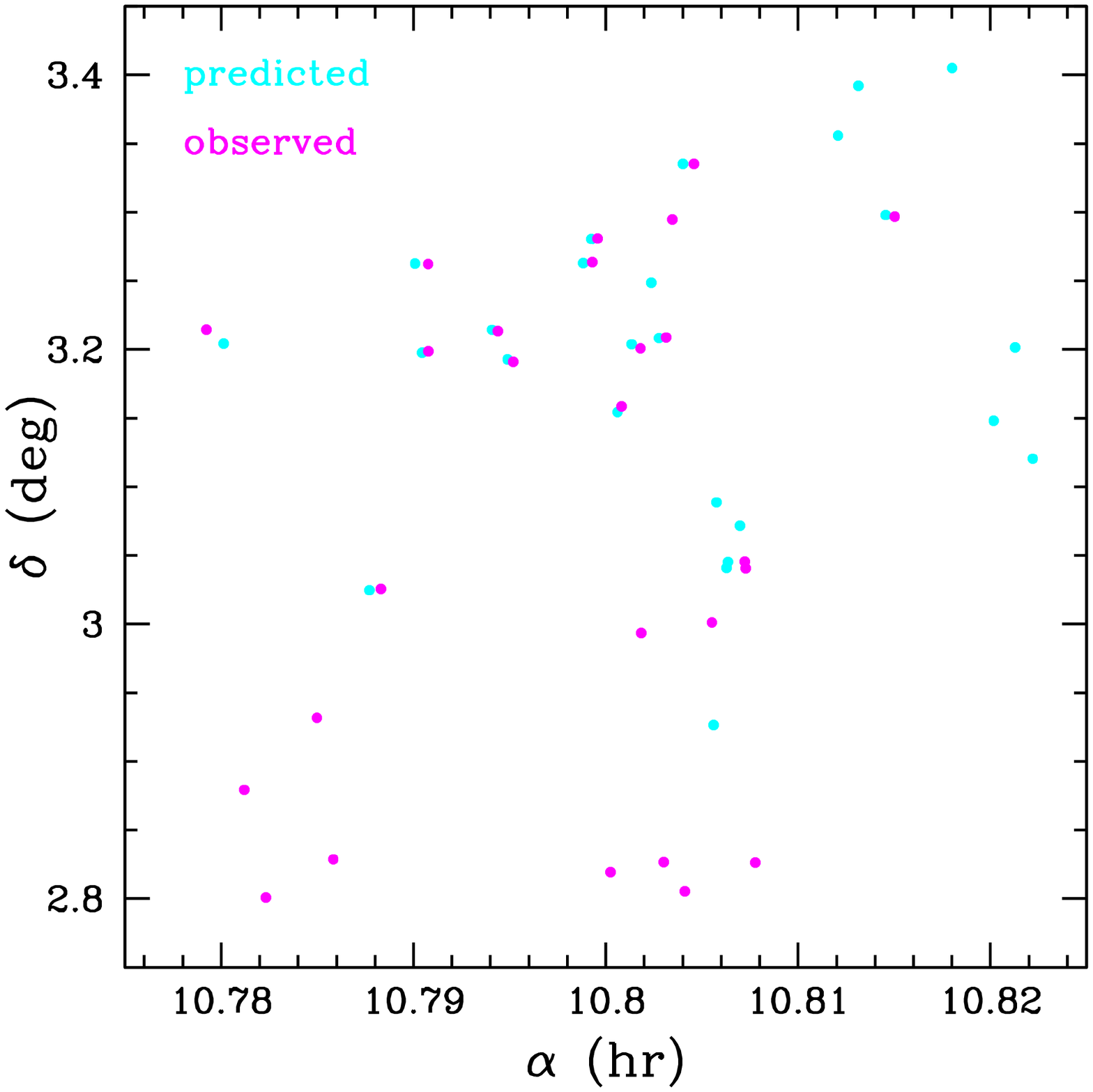}}
  }
  \caption{Pairing the unknown objects observed in multiple nights based on positions derived 
           from the extrapolated arc. Cyan points stand for extrapolated positions of objects observed 
           in the first night and magenta points mark objects observed in the second night in the 
           follow-up field. } 
  \label{fig1}
\end{figure*}

In order to secure some unknown objects, we observed at ESO/MPG nine fields in multiple nights, 
seven imaged during two nights and two fields during three nights. To identify objects observed 
in multiple nights, we extrapolated in time the arcs of all moving sources observed during the 
first night using a least square fitting code written by our team. 
By comparing the extrapolated positions of objects observed in one field during the 
first night for the observing date corresponding to the second night with the positions of the 
objects observed in the follow-up field during the second night, one could easily pair objects 
in the $\alpha-\delta$ position space. In case of crowding and close matches, the magnitude 
represents a second indicator. Using this pairing technique we matched all the unknown objects
observed during multiple nights, namely 43 objects observed during two nights and 8 objects 
observed during 3 nights. Thanks to this confirmation, most of these objects became credited 
discoveries. Figure~\ref{fig1} presents an example of such identification of objects observed 
in a crowded field ($9^{\circ}$ from opposition and $4^{\circ}$ south of the ecliptic).
With cyan circles we plot predicted positions and with magenta circles observed positions. 

The reduced images were analysed and measured using Astrometrica \citep{raa11}, carefully 
visually blinking all images of the same field in order to detect all moving objects. 
Around 100 reference stars (UCAC-2 for Swope data or USNO-B1 for ESO/MPG and INT data) were 
used for each CCD to perform the astrometry, using a linear model in the cases of WFI and 
Swope known to have small field distortion and fitting a 3-degree polynomial in the 
case of WFC to accommodate its larger optical field distortion. 
According to the literature and also to our derived astrometry (Section~\ref{NEA}), 
the astrometric errors due to the field distortion appear very small for the WFI field, being 
mostly within $0.1^{\prime\prime}$ across most of the WFI field (about $90\%$), 
according to the field distortion pattern derived by \cite{ass10} which shows the largest 
distortion in the corners of the mosaic (up to $0.53^{\prime\prime}$) and some distortion 
around $0.2^{\prime\prime}$ in the upper and lower edge and the centre of the mosaic. 
Relative photometry was derived by Astrometrica using from a dozen to a few hundred catalog 
stars visible in the field, the reduced asteroid magnitudes having an uncertainty of about 
0.1 mag. For all runs, we decided to use Astrometrica in preference to other software because 
of its simplicity, common platform and simple installation and usage by all members of the 
team, as many of the data reducers were students and amateurs. We inspected the data visually 
instead of using automated software, because of the relatively low volume of data per run, and 
mostly because the human eye and brain are known to detect faint moving sources better than 
the computer. We detected asteroids as low as ca $1.5\sigma$ level from noise, which allowed 
us to recover many faint targets inaccessible to other automated surveys. 

All the reduced data (output of Astrometrica in MPC format) were collected by the PI of the 
run, checked for errors using the EURONEAR O--C calculator or the FITSBLINK residual calculator 
\citep{skv10a}, then submitted to the MPC in three groups: the observed NEAs, the known asteroids, 
and the unknown asteroids (possibly discovered by us).

\section{Results}
\label{results}

During all three runs, we observed effectively in total about 11 nights, reporting positions 
and magnitudes for 100 program NEAs (7 VIs, 29 PHAs and 64 NEAs), 558 known MBAs and 628 unknown 
moving objects, in total 1286 objects observed in 29 square degrees total surveyed field. 

Table~\ref{table2} presents the overview of our observations at ESO/MPG, Swope and INT, 
listing the number of observed objects, number of reported positions and the standard deviation 
of the orbital fit using our data (according to NEODyS). We classify the objects in three groups: 
observed NEAs, known MBAs, and the unknown (unidentified) objects. We further classify the unknown objects
in three groups: official discoveries, later identifications and outstanding objects. 
We include the number of nights observed at each facility, the number of observed fields, the 
total sky coverage (in square degrees) and the limiting magnitude for each facility. Next, 
we give the total number of objects. Finally, we conclude with some density statistics 
which will be discussed in Section~\ref{stats}. 

\subsection{Program NEAs}
\label{NEA}

In the left panels of Figures~\ref{fig2}, \ref{fig3} and \ref{fig4} we plot the O--C residuals 
(observed minus calculated) in $\alpha$ and $\delta$ for the program NEAs observed at ESO/MPG, 
Swope and INT, based on the orbital fits available on 22 Nov 2010 \citep{neo11}. The O--C 
standard deviations are $0.15^{\prime\prime}$ for the ESO/MPG dataset, $0.39^{\prime\prime}$ 
for Swope and $0.42^{\prime\prime}$ for the INT. 

In total, 27 MPC and MPEC publications include data from our three runs. 
Table~\ref{table3} includes these references: 12 publications containing our Swope 
observations, 10 our ESO/MPG observations and 6 publications including our INT data. 

\citet{bir10} presented the most important NEAs recovered at ESO/MPG and Swope. 
At ESO/MPG we observed 12 faint objects (6 VIs, 3 PHAs and 3 NEAs) less than one month after 
discovery, and also recovered 2 faint objects (1 PHA and 1 NEA) at their second or later opposition.  
With Swope we observed 25 objects (all NEAs) soon after discovery, recovering also 2 objects 
(one NEA and one PHA) at their second or later opposition. From Table~\ref{table1} we can 
count the number of recoveries using the INT: 25 objects soon after discovery (1 VI, 8 PHAs and 
16 NEAs) and 6 faint objects having large uncertainty recovered at their second or later 
opposition (2 PHAs and 4 NEAs). 

Nine especially important NEAs were observed with the INT. We mark them with * in the first 
column of Table~\ref{table1} and we discuss them next. Thanks to the the large aperture of the 
INT and the large field of the WFC, we eliminated 3 NEA candidates and 1 VI. 
First classified as a NEA, 2009 CB2 had a poor orbit (2 day arc) and a very high sky uncertainty 
($3\sigma=1000^{\prime\prime}$). Thanks to the large field of WFC, we recovered this object one 
week later, allowing its orbit to be improved and eliminating it from the NEA list. 
A similar case was 2010 CF12, originally classified as a NEA based on a small 3 day arc. 
Although the MPC did not list its sky uncertainty by the time of our INT run, we recovered this 
object one week later and eliminated it from the NEA list. Another object degraded from the 
NEA class to the MBA class was 2010 DC. It had a small arc based on 4 nights data and a sky 
uncertainty of about $1^{\prime}$, allowing fast recovery two weeks later at the INT. One of 
the most important detections of the INT was 2009 VR25, a new object classified as a VI based on 
its original poor two night orbit. Although quite faint ($V=20.7$ mag) and having very large sky 
uncertainty ($3\sigma=1400^{\prime\prime}$), we could find it, enabling its reclassification
from the VI to NEA class. 

We recovered 3 NEAs at their second opposition with the INT. 
2007 RM133 was discovered in 2007, having a short arc (one month) at the time of 
the INT run. Despite its very faint magnitude ($V=21.4$ mag) and very large sky uncertainty 
($3\sigma=490^{\prime\prime}$), we recovered it 3 years later, allowing a substantial 
improvement of its orbit \citep{hol10}. 
Another second opposition recovery was 2003 SJ84, another NEA observed for only one month,
following its discovery in 2003. Although very faint ($V=21.4$ mag) and having quite large 
positional uncertainty ($3\sigma=81^{\prime\prime}$), we recovered it six years later 
about $21^{\prime}$ away from its predicted position (15 times more than its nominal
MPC $3\sigma$ value) and we improved its orbit \citep{fit09c}. 
Our best ever second opposition recovery was 2000 SV20. This NEA was observed for 3 
months following its discovery in 2000 and recovered 10 years later by the INT/WFC
about $7^{\prime}$ away from its predicted MPC position (7 times more than the nominal MPC
$3\sigma\sim1^{\prime}$ value).

Using the INT, in 2010 we followed two Arecibo radar targets. 2010 DJ1 was requested by 
NASA in Feb 2010 \citep{ben10}, having an uncertainty ($3\sigma=190^{\prime\prime}$) which 
allowed its INT recovery only three days after discovery. 2007 EF was another Arecibo target, 
moving very fast ($\mu=20~^{\prime\prime}$/min) but allowing successful WFC imaging at 
$V=18.2$ mag in only 10 sec exposures. 

\subsection{Known MBAs}
\label{MBA}

During all three runs in the observed program NEA fields, we observed incidentally 
a total of 558 known MBAs. 

\begin{figure*}
  \centerline{
    \mbox{\includegraphics[width=7.5cm]{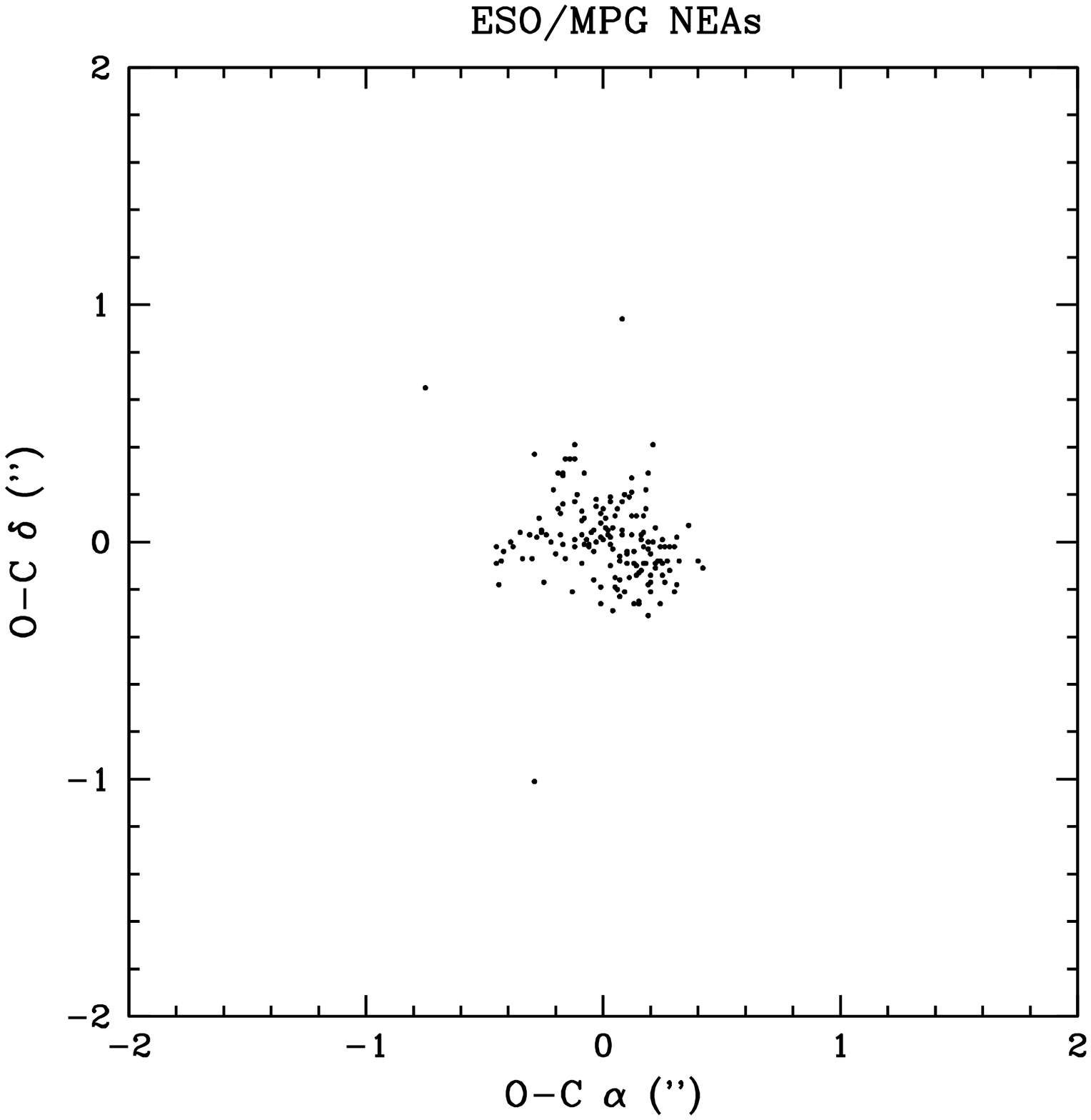}}
    \mbox{\includegraphics[width=7.5cm]{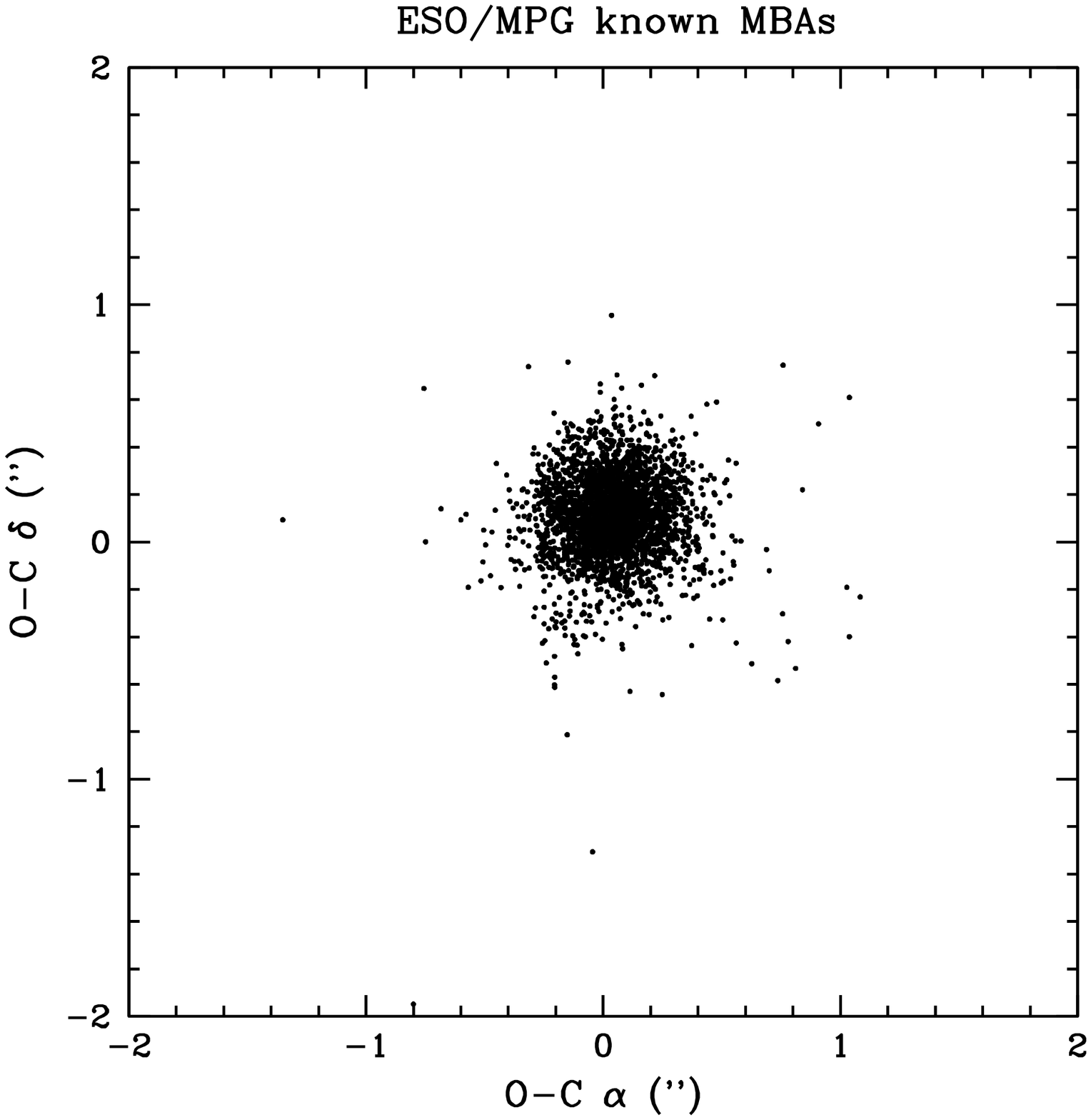}}
  }
  \caption{O--C (observed minus calculated) residuals for program NEAs and known MBAs observed at ESO/MPG. }
  \label{fig2}
\end{figure*}

\begin{figure*}
  \centerline{
    \mbox{\includegraphics[width=7.5cm]{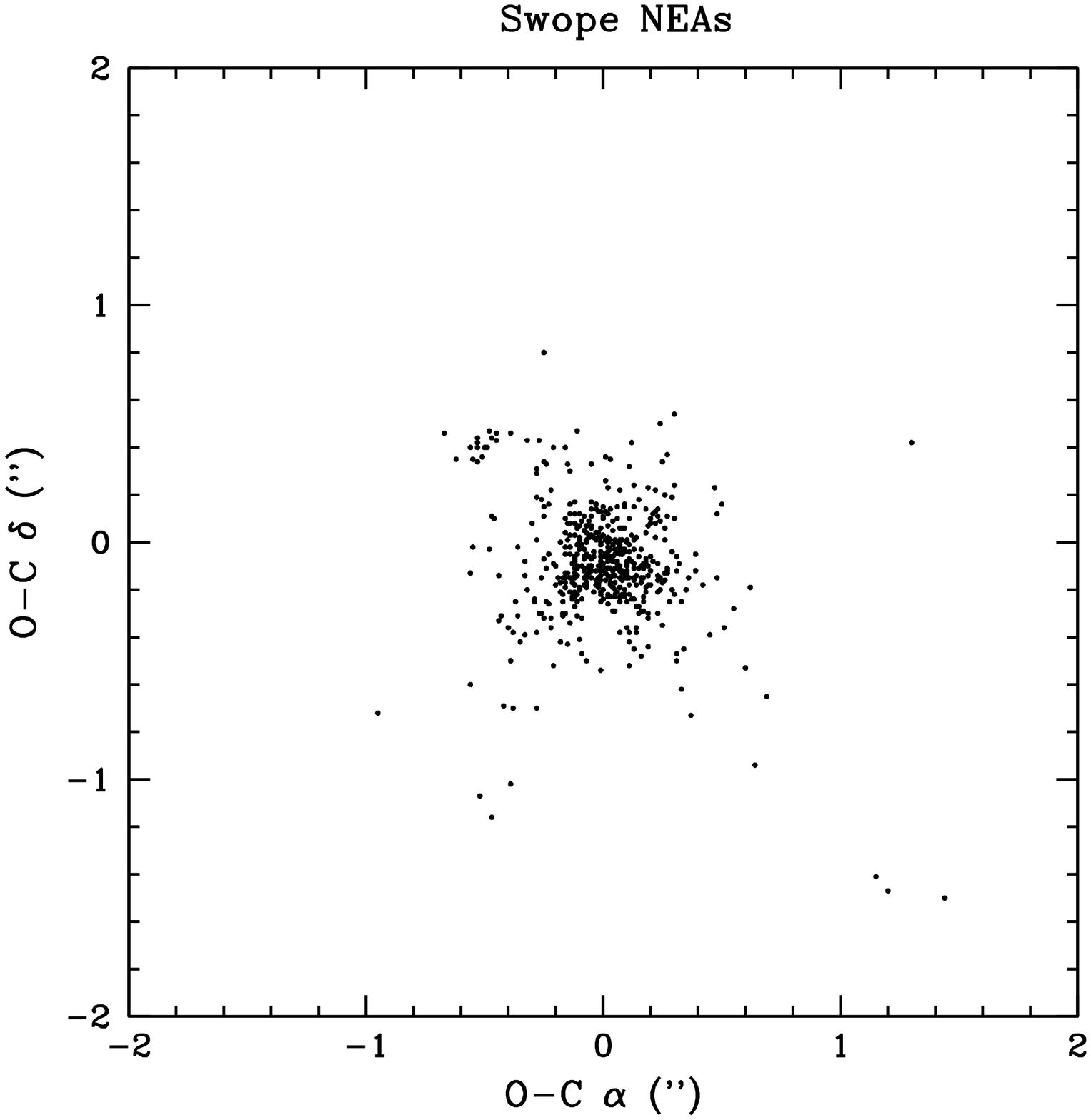}}
    \mbox{\includegraphics[width=7.5cm]{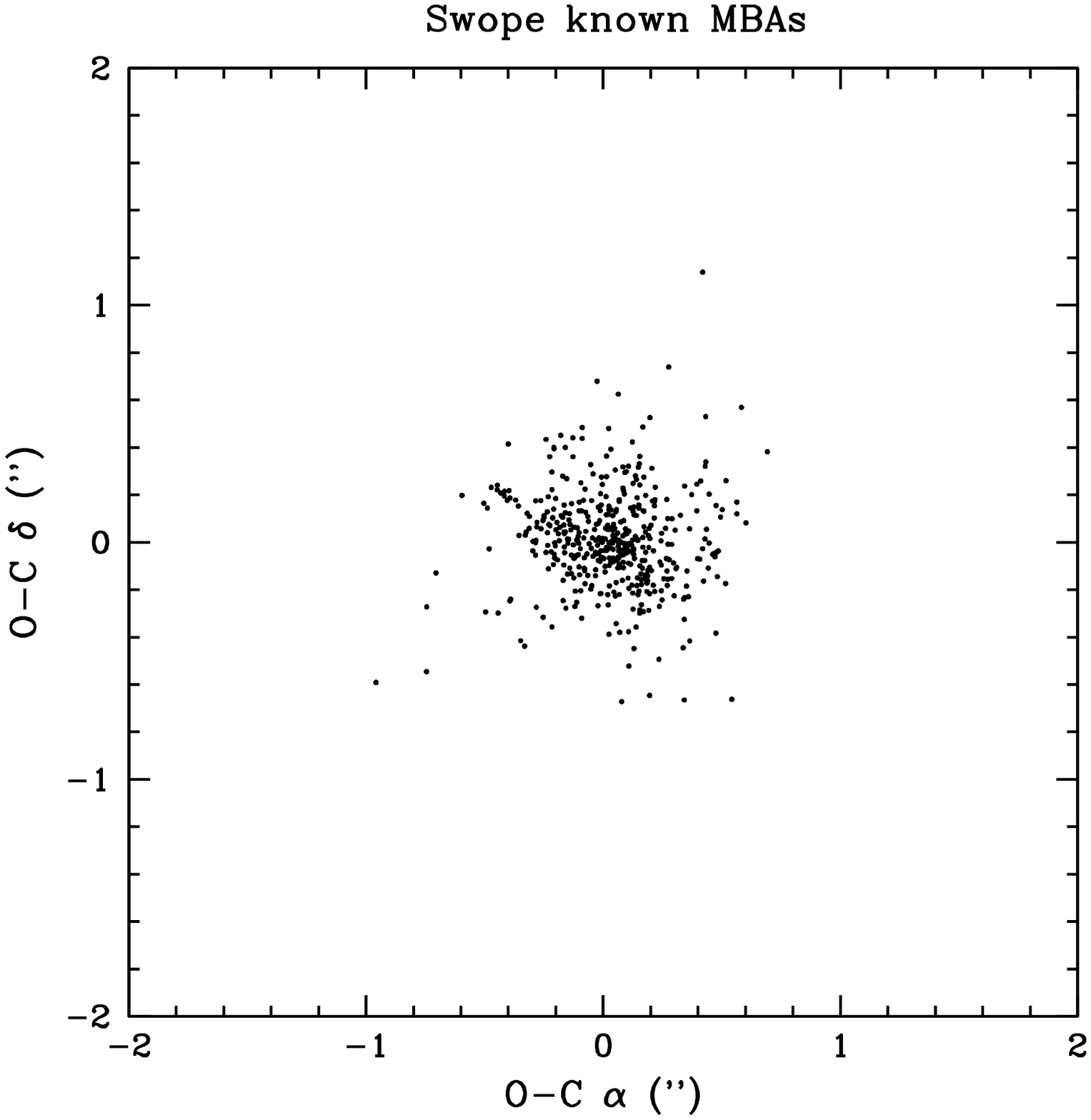}}
  }
  \caption{O--C (observed minus calculated) residuals for program NEAs and known MBAs observed with Swope telescope. }
  \label{fig3}
\end{figure*}

In the right panels of Figures~\ref{fig2}, \ref{fig3} and \ref{fig4} we plot the O--C 
residuals for the known MBAs observed in all three runs based on their current orbits 
available at 22 Nov 2010 \citep{ast11}. The standard deviations are $0.15^{\prime\prime}$ 
for the ESO/MPG dataset, $0.18^{\prime\prime}$ for Swope and $0.66^{\prime\prime}$ for 
the INT. 

The standard deviations of the ESO/MPG dataset for both NEAs and MBAs are very small,
proving the excellent quality of this telescope equipped with the WFI which allowed 
very accurate astrometry across its whole large field. Nevertheless, a systematic offset 
to the north of $0.111^{\prime\prime} \pm 0.001^{\prime\prime}$ in $\delta$ and 
$0.037^{\prime\prime} \pm 0.001^{\prime\prime}$ in $\alpha$ shows up in the right panel of 
Figure~\ref{fig2} for the known MBA dataset observed at ESO/MPG, with the observed median 
position located north-west of the calculated ephemerides. This effect is consistent with 
results of \citet{tho08a} who found a surprising systematic offset of the astrometry of the 
asteroid Apophis of about $0.2^{\prime\prime}$ based on 200 observations reduced with the 
USNO-B1 catalog (which was also used by us to reduce our ESO/MPG data). By comparing 
USNO-B1 with ICRF, the authors determined for the USNO-B1 an average declination offset 
of $+0.116^{\prime\prime}$, in perfect agreement with our findings. 

The standard deviation of the Swope datasets is $0.39^{\prime\prime}$ for the NEA data and 
$0.11^{\prime\prime}$ for the MBA data, and no sample shows any systematic offset with respect 
to the origin. We reduced Swope data with the UCAC-2 catalog, known to have better astrometry than 
USNO-B1. Both for Swope and ESO/MPG runs, the standard deviation for the MBA sample is smaller
than that of the NEA sample probably because of the higher S/N due to brighter and slower 
moving MBA objects, compared with the fainter and faster moving program NEAs. 

The INT/WFC shows relatively large residuals for both NEA and MBA samples (up to 
about $2^{\prime\prime}$ and standard deviation $0.66^{\prime\prime}$ as quoted above)
and only for this facility the deviation of the MBA sample is larger than that of the 
program NEAs. 
Bad weather did not impede the astrometry for our INT run, catalog stars being quite 
bright in a 2m telescope resulting in good S/N and sub-arcsec stellar positions. The USNO-B1 
catalog has an average astrometric accuracy of $0.2^{\prime\prime}$, too small to explain 
the larger residuals of the INT astrometry. 

The larger residuals for known MBAs than for program NEAs observed with INT/WFC 
can however be explained taking into account that MBAs were imaged across the entire WFC 
field which is heavily affected by the field distortion at the INT prime focus, compared 
with the central CCD\#4 where most of the NEAs were observed and the WFC field has the 
smallest distortion. Given this, for any future INT/WFC work we plan to correct the 
image field before data reduction. 

\subsection{Unknown Objects}
\label{unknown}

A total of 467 unknown moving objects were identified at ESO/MPG, 41 at Swope and 120 at 
the INT (Table~\ref{table2}). The observed proper motion 
for most objects was compatible with MBAs with a few exceptions
discussed next. 

Throughout our paper and in Table~\ref{table2} we classify the unknown objects in three 
categories: {\it official discoveries} (confirmed by the MPC according to their DISCSTATUS 
monthly list), {\it later identifications} (unknown objects which could be linked with 
existing arcs) and {\it outstanding objects} (waiting for orbital links from independent 
observations, possibly to be credited to us later). 

\subsubsection{Unknown MBAs}

We include in the Appendix A (available only in the online version of our paper) seven 
tables listing all the unknown objects observed at ESO/MPG (Tables A.1, A.2 and A.3), 
Swope (Tables A.4 and A.5) and the INT (Tables A.6 and A.7). We give first the official 
discoveries, then later identifications and finally outstanding objects. 

Table A.1 includes our official discovered asteroids at ESO/MPG. We give first 
the EURONEAR object acronym (based on the initials of the surname of the observers and 
reducers\footnote{VB = Vaduvescu and Birlan; TU = Tudorica; SO = Sonka; OP = Opriseanu; 
VI = Vidican; TO = Toma; VA = Vancea.}), then the official designation (from the MPC), three
main orbital parameters (semimajor axis $a$, eccentricity $e$ and inclination $i$), Earth 
minimum orbital intersection distance $MOID$, absolute magnitude $H$, number of observed 
positions, arc length (in days or years), rms of the O--C residuals for the orbital 
fit $\sigma$, observed apparent magnitude $R$, ecliptic latitude $\beta$, Solar elongation 
$\epsilon$ (both in degrees) and apparent proper motion $\mu$ (in 
$^{\prime\prime}$/min). We distinguish directions east and west of
opposition by letting $\epsilon$ increase above $180^{\circ}$ for
fields to the east. 

For each object we give in the first line the orbital elements fitted with the FIND\_ORB 
software \citep{gra11a} based on our observations only and for the standard epoch 
$MJD = 54520.0$. 
On the second line we list the orbital elements taken from the MPC database \citep{mpc11} 
calculated by fitting all MPC available observations for an epoch close to the mid-point 
interval of those observations. 

Table A.2 includes the unknown objects (at the date of the run) which were identified 
later (in Nov 2010) with known objects, based on the checks of the MPC database and the 
MPC automatically assigning designations. Our calculated orbits, presented again in the 
first line, are based on the very short available arc (observations acquired in less than 
one hour), and so should be regarded with caution.

For both official discoveries and later identified objects, the orbital elements from 
the first line are very close to the official elements in the second line. Although most 
of our fits are based on very short arcs observed during 1--3 nights at ESO/MPG and only 
one night at Swope and INT, one can observe the success of the FIND\_ORB fit for most 
objects, especially for the $a$, $e$ and $i$ parameters which will be used in statistics 
later. In particular, based on 116 paired orbits available from the
three runs, we can compare FIND\_ORB versus MPC by calculating the median
values of the differences in $a$, $e$, $i$ and $H$, which are
$0.20$ AU, $0.06$, $1.05^{\circ}$ and 0.70 mag, respectively. 

Table A.3 includes the remaining unknown objects observed at ESO/MPG which could not be 
identified according to the present MPC database (Nov 2010). For them we give only our 
calculated orbits, based on very short arcs observed only in one night. The exact elements should
be regarded with caution but are usable for statistics in Section~\ref{disc}. 

Table A.4 lists the later identified objects observed with the Swope telescope, while 
Table A.5 gives the outstanding unknown objects observed with Swope. Similarly,
Tables A.6 and A.7 give the later identified objects and outstanding unknown objects
observed at the INT. 

According to the MPC (Jan 2011 DISCSTATUS), our ESO/MPG run produced 58 official discoveries.  
Most of the people involved in EURONEAR work on a voluntary basis and include students 
and amateur astronomers based in Romania. Indeed, the entire ESO/MPG team included people 
of Romanian origin who became the first Romanian discoverers of minor planets \citep{vad09}. 
Given these, in 22 December 2008 we proposed to the Working Group for Small Body Nomenclature 
(CSBN) of the International Astronomical Union (IAU) a list including 12 Romanian names. 
In Jan 2011 the first two of our discovered asteroids received numbers, namely 
(257005) = 2008 EW152 = VBTU207 (our acronym) and (263516) = 2008 EW144 = VBTU224, being 
eligible for naming. 

\subsubsection{NEO Candidates}

In this section we will use three tools to check all our unknown objects for potential 
Near Earth Objects (NEOs). In Table~\ref{table4} we include all NEO candidates derived 
from all three methods, marking in bold the best candidates. Besides the observed 
properties (quantities $\mu$, $\epsilon$, $R$, number of positions and O--C standard 
deviation), we include in this table the orbital parameters ($a$, $e$, $i$) and data 
derived from the three methods (MOID in column 4, MPC score in column 10 and the 
Model in the last column). 

For the first method we plot in Figure~\ref{fig5} the apparent proper motion $\mu$ 
versus the Solar elongation $\epsilon$ for all the unknown objects observed at ESO/MPG 
(red), Swope (green) and the INT (blue). Observed in a given field near opposition 
(close to $\epsilon\sim180^{\circ}$), MBAs are expected to show proper motions 
distributed in a small vertical ``finger''-shaped region with proper motions between 
$\mu\sim0.3-0.7~^{\prime\prime}$/min, depending on their location in the main belt. 
Observed further away from opposition, MBAs should show smaller proper motions owing to 
the larger distance and velocity projection effect. Both $\mu$ and $\epsilon$ represent 
quantities measured directly from observations, not being affected by the uncertain 
orbits derived from the short arc. Therefore the $\mu-\epsilon$ plot represents an 
important method to search a survey for fast moving objects including NEAs, PHAs and 
other NEOs. 

Let us consider the very basic orbital model which assumes the (prograde)
asteroid orbit circular and coplanar with the circular orbit of Earth, and
the asteroid at least $90^{\circ}$ away from the Sun. Following \citet{kol99}
we express the asteroid proper motion as a function $\epsilon$.  Let $\Delta$
be the angle between the directions of Sun and asteroid as seen from Earth
($\Delta$ is $\epsilon$ or $360^\circ-\epsilon$, with
$0 < \Delta < 180^\circ$).  Let $v_E$ and $v_a$ be the orbital velocities of
Earth and asteroid, $\phi$ the angle as seen from the asteroid between the
direction of Earth and that of the asteroid's orbital motion, and $E$ the
difference $180^\circ$ minus the angle (as seen from Earth) between the
direction to the asteroid and the direction of Earth's orbital motion (so $E
= \Delta - 90^\circ$). Then the angular speed of the asteroid $\omega$ as
seen from Earth is the difference in the projected velocities perpendicular
to the Earth--asteroid direction, divided by the Earth--asteroid distance:

\begin{equation}
\label{eq1}
\omega = \frac{v_a \sin \phi - v_E \sin E} {d}
\end{equation}
\noindent
From the sine law applied to the triangle Sun--Earth--asteroid:

\begin{equation}
\sin \phi = \sqrt {1 - \frac{a_E^2 \sin^2 \Delta}{a_a^2} }
\end{equation}
\noindent
The cosine rule applied to the same triangle, solving the resulting quadratic
for $d$, gives:

\begin{equation}
d = a_E \cos \Delta + \sqrt {a_a^2 - a_E^2 \sin^2 \Delta}
\end{equation}
\noindent
where we dropped the minus solution because $d$ should always be positive. 
Kepler's third law (assuming Earth and asteroid masses very small compared 
with Sun's mass) implies $v_a = \sqrt{\frac{G M_S}{a_a}}$ and $v_E = \sqrt{\frac{G M_S}{a_E}}$
where $G = 6.673 \times 10^{-11}$ m$^3$ kg$^{-1}$s$^{-2}$ is the gravitational constant, 
$M_S = 1.989 \times 10^{30}$ kg the Sun's mass and $a_E=1.496 \times 10^{11}$ m
the Earth's semimajor axis. \\

\noindent
Substituting all these terms in Equation~\ref{eq1}, and since
$\sin E = -\cos \Delta$, we obtain the following formula for the apparent
angular speed of the asteroid as a function of $\Delta$ and $a_a$:

\begin{equation}
\omega = \frac{\displaystyle
	\sqrt{\frac{G M_S}{a_a}} \sqrt {1 - \frac{a_E^2 \sin^2 \Delta}{a_a^2}} + 
        \sqrt{\frac{G M_S}{a_E}} \cos\Delta}{a_E \cos \Delta + \sqrt {a_a^2 - a_E^2 \sin^2 \Delta}} 
\end{equation}
\\
\noindent
Finally, the proper motion of the asteroid $\mu$ in arcsec per minute
can be calculated from the angular speed $\omega$ in radians per second:

\begin{equation}
\label{eq2}
\mu = \omega \times \frac{180\times3600}{\pi} \times 60
\end{equation}

We can use Equation~\ref{eq2} and the asteroid semimajor axis $a_a$ as a model to 
map the expected limits for the proper motion of MBAs (defined between $a_a=2.0$ 
and $a_a=3.5$ AU). In the context of this basic orbital model where we consider
proper motions (as a function of elongation) only, we may also represent NEOs by
considering orbits with $a_a < 1.3$ AU. By plotting 
the values of $\mu$ between solar elongations $90^\circ$ and $270^\circ$ (e.g., using one 
degree step in $\epsilon$), one can draw the limits corresponding to these populations. 
In Figure~\ref{fig5} we plot with dotted magenta lines the curves corresponding 
to these limits. The curves are symmetric about 180$^\circ$ (in fact symmetry properties
of sky motions hold more generally in the non-coplanar case; see section 2.1 of \citet{jed96}. 

According to Figure~\ref{fig5}, most of the 628 unknown objects agree with our model,
in the sense that most are consistent with asteroids from the main belt. They are located 
at the bottom of the plot around $\mu=0.2-0.8~^{\prime\prime}/$min and between the two 
dotted curves corresponding to $a_a=2.0$ and $a_a=3.5$ AU. 
About 16 objects ($2.5\%$ of the total) marked with circles rise above the $a=1.3$ NEO limit 
and above the main vertical group at the respective elongation.
We mark these objects with ``fit'' or ``best'' 
in the last column of Table~\ref{table4}, treating them as potential NEOs. One can 
clearly distinguish three major outliers showing the fastest 
proper motions, namely VBVI213 at $\mu=7.17~^{\prime\prime}/$min and VBTU222 at 
$\mu=2.12~^{\prime\prime}/$min observed at ESO/MPG, and VTU021 at 
$\mu=4.61~^{\prime\prime}/$min observed at the INT. We mark them with ``best'' in the 
last column of Table~\ref{table4}, treating them as our best NEO candidates. 

The fastest object was recorded under the acronym VBVI213 and it moved about 10 times 
faster than all other MBAs observable close to opposition, so it represents our best NEO 
candidate. This object was clearly visible on 8 CCD images, leaving a 20 pixel trail 
owing to the relatively long 2 minute exposures. It moved in the opposite direction 
and about 10 times faster than all other asteroids visible in the same field. In 
Figure~\ref{fig6} we include the image of this field (CCD\#5 of WFI), presenting the 
corresponding 8 frame animation in the online electronic version of the paper. The image 
is displayed in normal sky orientation and the field of view is about $8^\prime \times 16^\prime$ 
(one WFI CCD), with pixel size $0.714^{\prime\prime}$ (in $3\times3$ binning). Four MBAs marked 
with circles are visible moving to the upper right, while the NEA candidate is visible as 
a trail in the bottom part (enlarged twice in the left corner inset).
The exposure time was 2 minutes and the cadence 
between frames was 3.3 minutes. 

Most of our fields were observed near opposition
($150^\circ < \epsilon < 210^\circ$) and all agree well with our model. 
In Figure~\ref{fig5} there are about three fields located farther from 
opposition between $130^{\circ} < \epsilon < 140^{\circ}$ for which most unknown objects 
do not match our model which does not hold due to our basic (circular and coplanar) 
orbital assumptions (proper motions in the model are close to zero, whereas the real
orbits give a small but noticeable component to the proper motions).

Our second NEO search method uses the ``NEO Rating'' tool developed by the Minor Planet 
Center (MPC) which calculates a score for possible NEOs based on the 
expected proper motion of the MBA population distribution \citep{mpc11}. 
Running this tool for all unknown ESO/MPG objects, we obtained NEO scores (``No-ID'' 
probabilities) of $100\%$ for three objects: VBTU197, VBTU222 and VBVI213 which 
confirm our findings using the first model. Running the NEO Rating for all unknown 
INT objects we confirm with $100\%$ score two INT objects, VTU021 and VTD003, plus 
VITB01 with a relatively high score ($77\%$). We plot with circles all these NEO 
candidates in Figure~\ref{fig5} and we include the scores in Table~\ref{table4}. All 
other objects from all runs received very low MPC rates (smaller than $5\%$), 
consistent with our MBA classification derived from our model. 

Our third NEO search method uses the calculated MOID derived from the preliminary orbits 
derived with FIND\_ORB. The results are included in Table~\ref{table4} and they mostly 
agree with the other two methods, showing in most cases the reliability of the derived 
preliminary orbits, especially for the objects observed closed to opposition. 

Unfortunately most NEO candidates were observed in only one night and only one object 
was reobserved during a second night, namely VBTU203 = 2008 EN144. Two NEA candidates 
observed at INT were identified later with known objects, namely VTD003 = 2010 CJ33 and 
VIT005 = 2000 SN27. Based on their updated observed arcs, they are not NEAs, so we mark 
them by * in Table~\ref{table4}. Dropping them from the list, we count in total 
16 NEO candidates; this number should be considered an upper bound.
Four objects have MPC scores of $100\%$, small MOIDs (less than 0.1 AU) and they agree 
mostly with our model, so they represent our best 4 NEO candidates: VBVI213, VBTU222, 
VBTU197 (observed at ESO/MPG) and VTU021 (observed at the INT). We write their acronyms 
in bold in the first column of Table~\ref{table4}. There was no NEO candidate observed 
with Swope, checking all three methods. 

We checked the remote possibility that some NEO candidates could be identified with Earth 
artificial satellites or space debris. In this sense, we checked all our observed fields 
against known Earth satellites by using the satellite identification server developed by 
\citet{skv10b} based on the software SAT\_ID of \citet{gra11b}. No satellite with proper 
motion slower than $0.25^\prime$/min (corresponding to geostationary orbits) was found 
within one degree of any observed fields. Known and unknown space debris could also be
studied statistically, according to \citet{sch07}. Compared with NEO candidates, 
space debris move at very fast speed with angular velocities ranging from a few arc seconds 
per second (i.e., a few arcminutes per minute, at least 10 times faster than our fastest 
NEO candidate) to more than 1,000 arcseconds per second with respect to the stellar 
background. Thus, we drop any possibility that any of our unknown objects could be 
associated with artificial satellites or space debris. 

\section{Discussion}
\label{disc}

\subsection{Comparison with the known asteroid population}

We compare the major orbital parameters of all unknown objects observed at ESO/MPG, Swope 
and INT with the entire known asteroid population at 10 Dec 2010 (541,260 orbits) based on 
the ASTORB database \citep{bow11}. 

In Figure~\ref{fig7} we plot two classic asteroid orbital distributions,
namely $e$ versus $a$ (left) and $i$ versus $a$ (right). 
We overlay in colours all the unknown objects observed in our survey, 
including official discoveries, later identifications (based on MPC orbits) and outstanding 
objects (based on FIND\_ORB orbits). One can easily observe that the majority of our objects 
fit both orbital distributions very well, marking the four major Kirkwood gaps and a few 
known families. 
Only 105 points represent our official discoveries and later identifications, while 
523 points represent outstanding objects (five times more). Because outstanding objects have 
orbits calculated with FIND\_ORB, at least statistically this confirms the ability of 
FIND\_ORB to calculate preliminary orbits based on very small arcs. 

\subsection{Comparison between facilities}

In Figure~\ref{fig8} we plot the distribution of the observed apparent magnitude 
$R$ (left panel) and the calculated absolute magnitude $H$ versus semimajor axis $a$ 
for all unknown asteroids observed with ESO/MPG, Swope and INT.  The limiting
magnitude of each system is evident by the levels in the $R$ plot above which
the regions become depleted of data.  For each fixed limiting $R$, we expect
a negative trend of $H$ versus $a$, as seen in the right panel. 
We observed unknown MBAs up to $a\sim3.3$ AU, which is considered about half the 
outer main belt region \citep{yos07}. As expected, both 2m facilities sampled well 
the middle region ($2.6 \le a \le 3.0$) and the first half of the outer region 
($3.0 \le a \le 3.3$), while
well over half of unknown objects sampled with the Swope 1m are
in the inner region of the main belt ($2.0 \le a \le 2.6$). Moreover, according to 
Table~\ref{table2}, both ESO/MPG and INT discovered about the same number of MBAs as 
the number of known MBAs. Thus, a 2m survey could bring an important contribution to 
knowledge of the main belt, being expected to double the present number of known MBAs 
to more than one million. 

\begin{figure*}
  \centerline{
    \mbox{\includegraphics[width=7.5cm]{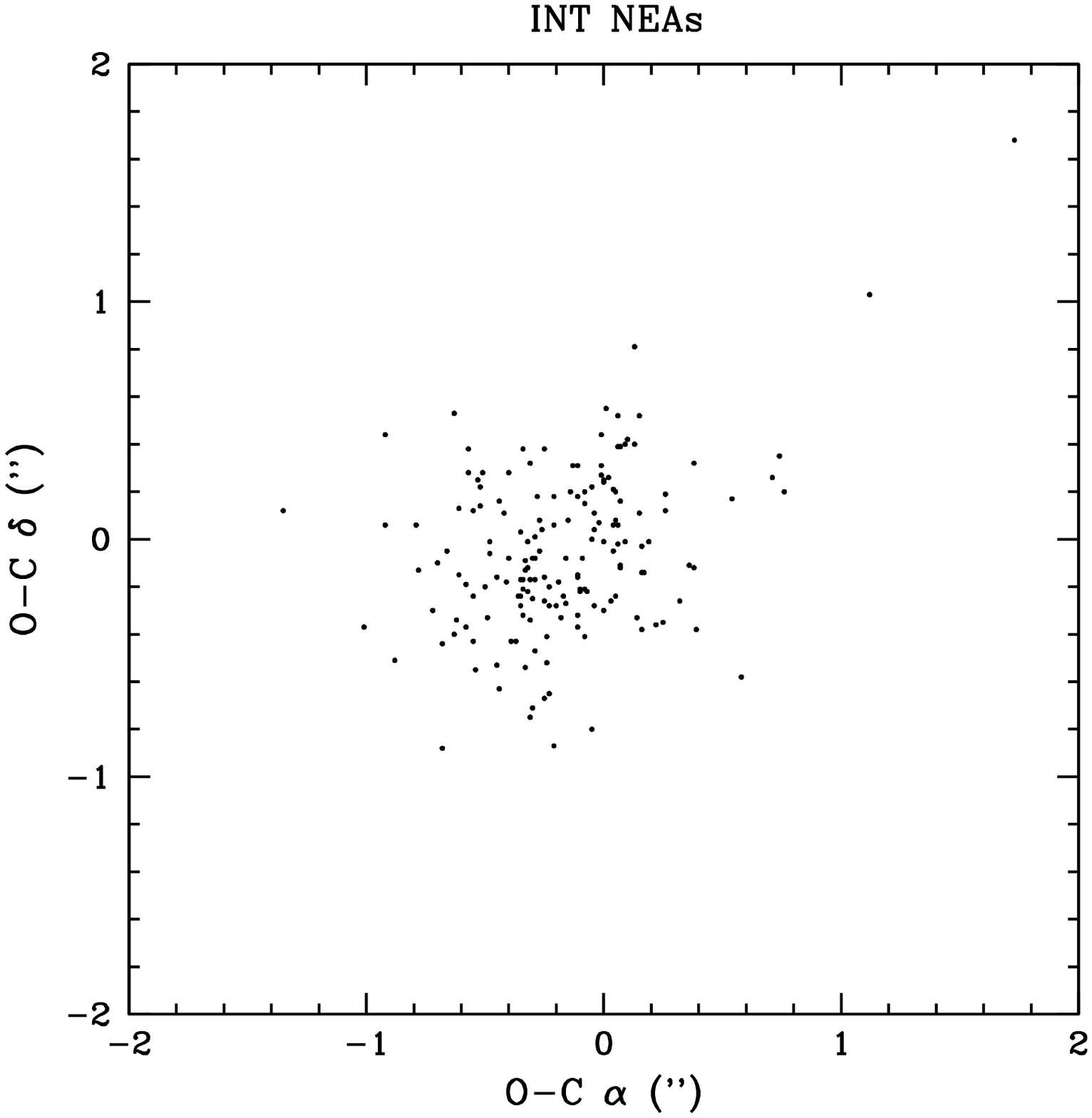}}
    \mbox{\includegraphics[width=7.5cm]{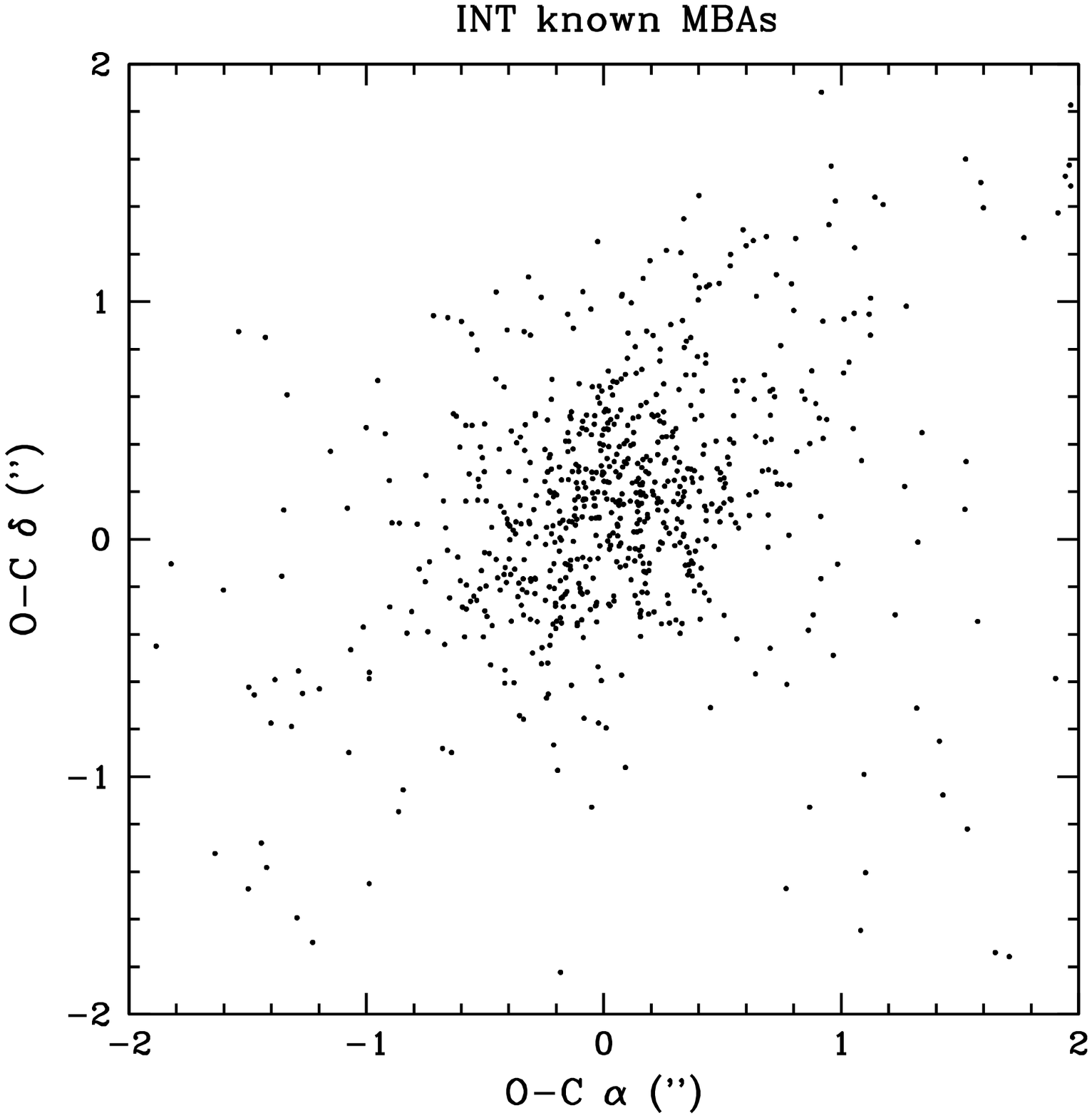}}
  }
  \caption{O--C (observed minus calculated) residuals for program NEAs and known MBAs observed with the INT. }
  \label{fig4}
\end{figure*}

\begin{figure*}
  \centerline{
    \mbox{\includegraphics[width=12cm]{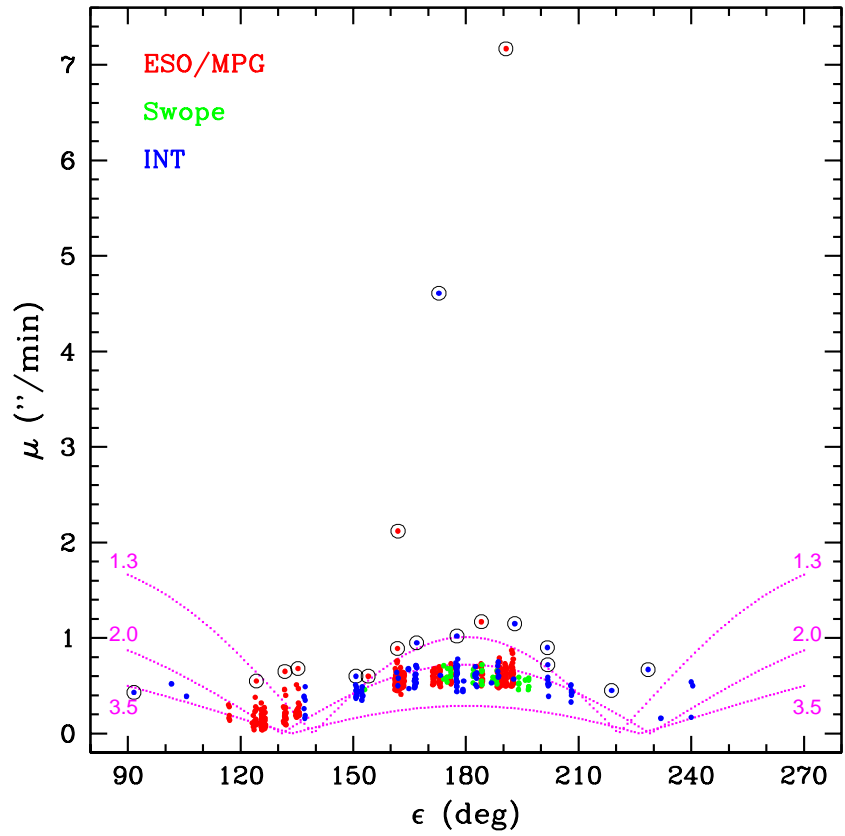}}
  }
  \caption{Basic orbital model using the asteroid observed proper motion $\mu$ and the Solar elongation 
           $\epsilon$. We plot all unknown objects observed at ESO/MPG (red), Swope (green) and INT (blue). 
           The three overlaid dotted magenta curves correspond to asteroids orbiting between $a=2.0$ and 
           $a=3.5$ AU (Main Belt) and $a=1.3$ (Near Earth Objects limit). The model allows us to easily flag NEO 
           candidates in a survey. We mark with circles our NEO candidates and we include their properties in 
           Table~\ref{table4}. } 
  \label{fig5}
\end{figure*}

\begin{figure*}
  \centerline{
    \mbox{\includegraphics[width=7.5cm]{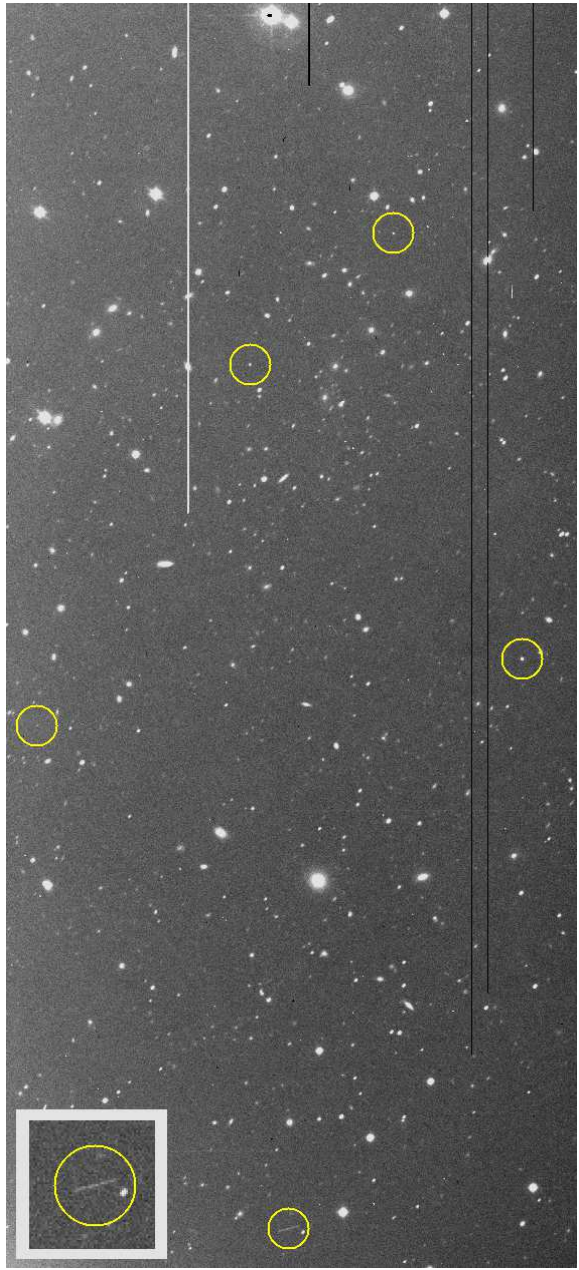}}
  }
  \caption{VBVI213, our best NEO candidate, at the bottom of the $8^\prime \times 16^\prime$ ESO/MPG 
	   WFI CCD\#5, moving in the opposite direction to the four MBAs marked above, and about 10 times faster.
           The image is displayed in normal sky orientation (N up, 
           E left) and the inset zooms in on the NEO candidate. An animation including all 8 available frames 
           is available online. } 
  \label{fig6}
\end{figure*}

\begin{figure*}
  \centerline{
    \mbox{\includegraphics[width=7cm]{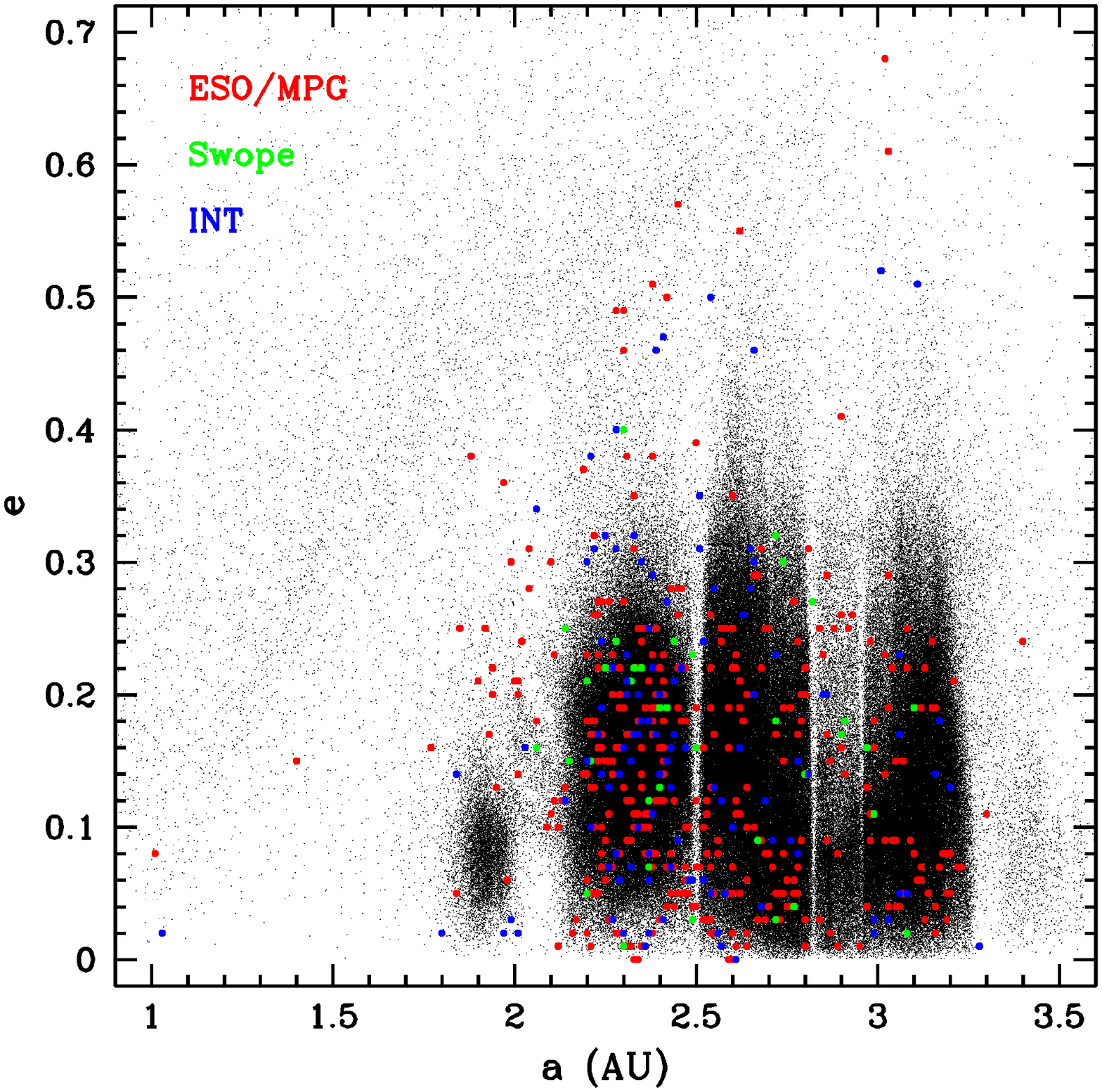}}
    \mbox{\includegraphics[width=7cm]{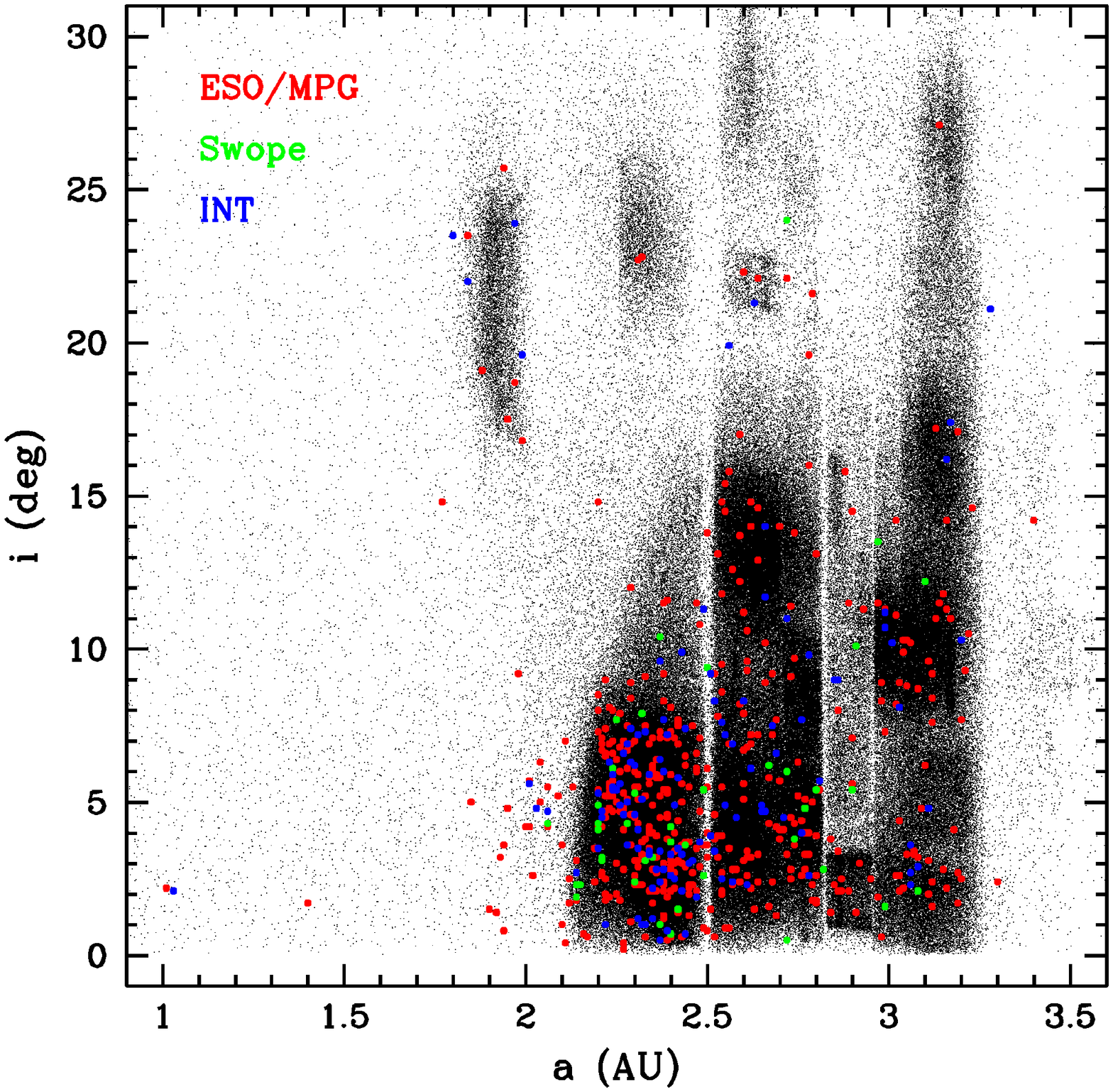}}
  }
  \caption{Orbital distributions of 628 unknown objects observed at ESO/MPG (red points), Swope (green) 
           and INT (blue) compared with the entire known asteroid population (ASTORB - 541,260 fine black points). 
           Although our preliminary orbits were derived using mostly short arcs, the distributions are consistent 
           with the known MBA population, showing the usefulness of the FIND\_ORB orbital fit in $a$, $e$ and $i$. } 
  \label{fig7}
\end{figure*}

According to the O--C plots for known MBAs (right panel of Figure~\ref{fig2}, \ref{fig3} and 
\ref{fig4}) and to the O--C standard deviation of $0.15^{\prime\prime}$ for both NEA and
known MBA datasets, ESO/MPG appears to have the best astrometry required for accurate follow-up, 
recovery and discovery across the whole WFI camera. With standard deviations of $0.18^{\prime\prime}$ 
for the known MBA dataset and $0.39^{\prime\prime}$ for the NEAs, the Swope telescope represents 
an adequate 1m facility for asteroid studies at limiting magnitude $R\sim20.5$ mag. The field of 
the INT/WFC appears the most distorted, these O--C positions showing the widest spread in 
Figure~\ref{fig4} and a standard deviation about 4 times larger compared with the other two 
facilities. Although the INT astrometry is acceptable around the centre of WFC and could be 
used to follow-up known objects expected to appear close to the centre, the INT field should 
be corrected in order to reach more accurate astrometry across the entire WFC field. 

Comparing the position of the centroid of the known bulk of MBAs with respect to the calculated 
positions, we conclude that the USNO-B1 catalog is less appropriate for astrometric reduction due 
to larger residuals, and we recommend instead UCAC-2 or UCAC-3 which appear to give more accurate 
results and therefore to be the current best representation of the Hipparcos
frame up to magnitude 16.  Nevertheless despite the superior accuracy from
using the UCAC system, it still has shortcomings for the astrometry of fast
moving objects such as the limited north declination coverage of UCAC-2 and
the faulty northern proper motion system of UCAC-3.

\subsection{Distribution of the unknown MBA and NEO candidates}

With the original 1m Spaceguard survey approaching its goal and limits, new 2m surveys such as 
Pan-STARRS will soon take over, increasing the detection limits in both size and depth in the 
Solar System. Based on our ESO/MPG and INT data, we briefly evaluate here the limits of such a 
2m survey. 

The left panel of Figure~\ref{fig9} plots the histogram showing the observed apparent magnitude 
$R$ for all the unknown objects observed at ESO/MPG (red colour), Swope (green), the INT (blue) 
and the total (black dots). Apparently, the dark time at ESO was most efficient to detect 
unknown objects at $R\sim20.6$ mag, allowing a limit $R\sim21.5$ mag. This detection limit is 
consistent with the actual 1.8m Pan-STARRS 1 which is expected to reach $R\sim22$ mag \citep{grav09}. 
The INT reached maximum detection at $R\sim20.2$ mag and a more shallow cutoff at $R\sim21.2$ mag, 
about $0.3$ mag less faint than ESO/MPG despite its larger aperture, probably due to the worse 
observing conditions at the INT. The 1m Swope reached a maximum detection at $R\sim19.8$ mag 
and a limiting magnitude at $R\sim20.4$ mag, about one magnitude lower than other 2m facilities. 
We include these limits in Table~\ref{table2}. 

\begin{figure*}
  \centerline{
    \mbox{\includegraphics[width=7cm]{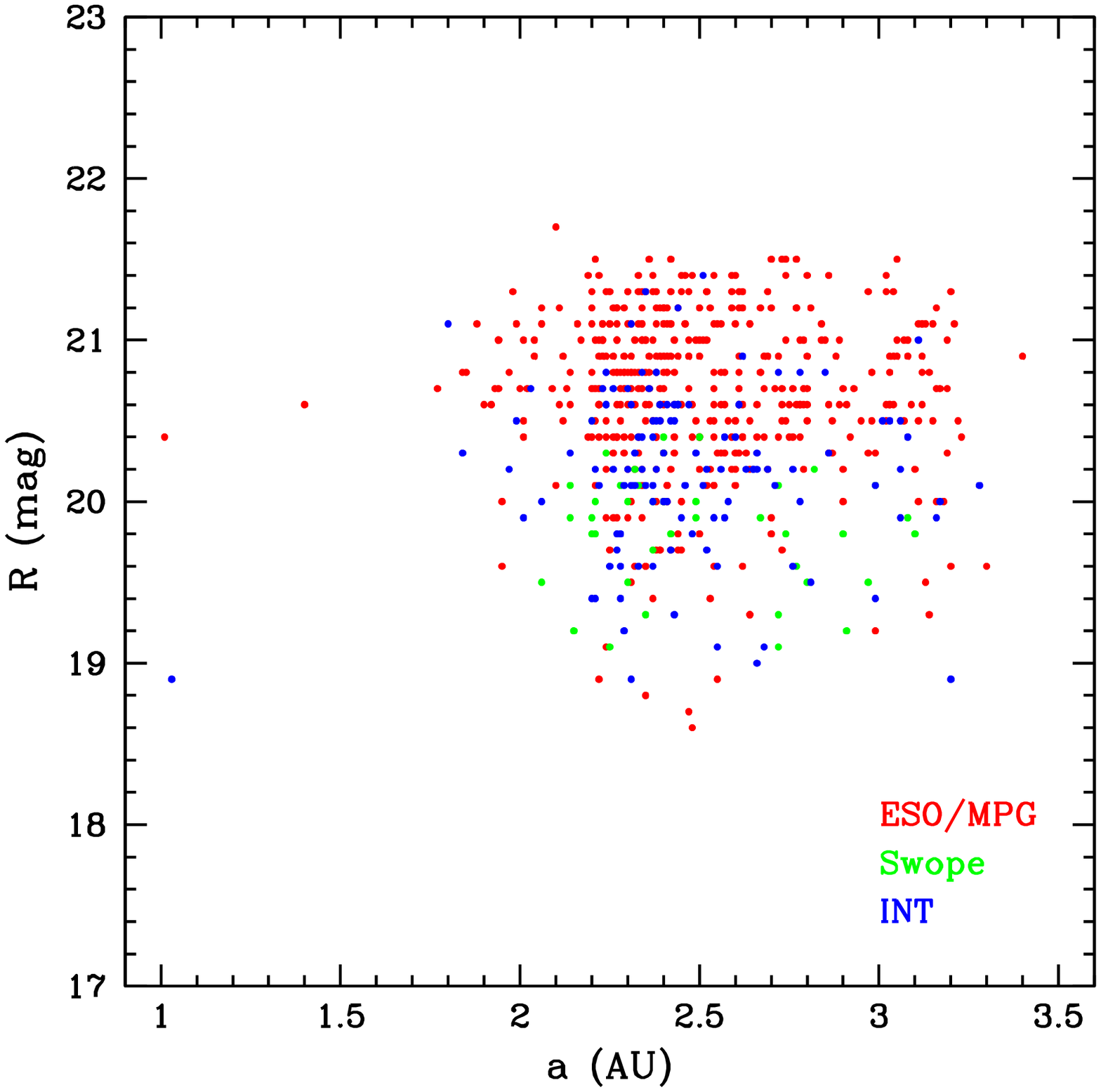}}
    \mbox{\includegraphics[width=7cm]{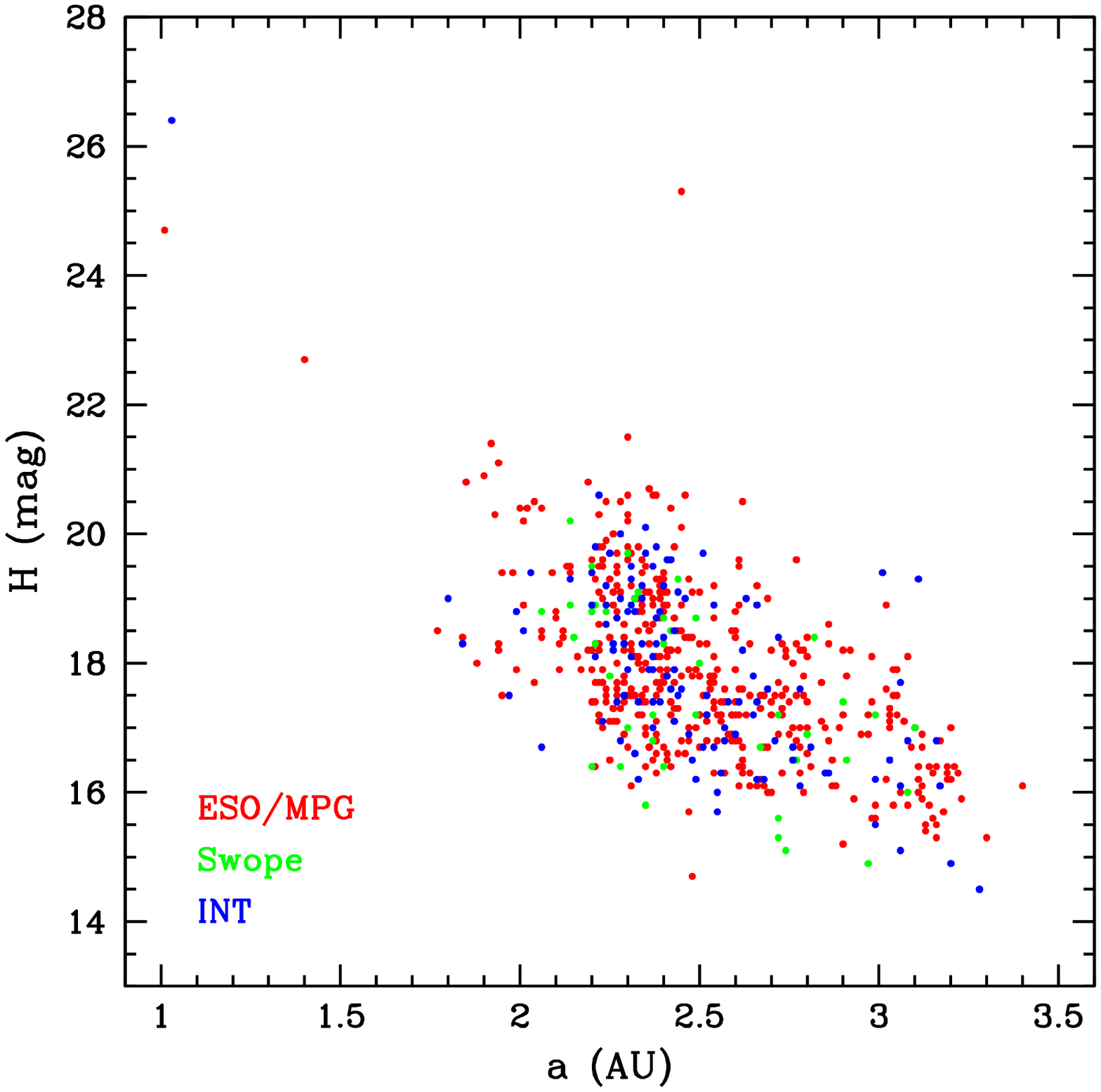}}
  }
  \caption{The observed apparent $R$ magnitude (left) and calculated absolute magnitude $H$ (right) versus 
          the semimajor axis $a$ for the ESO/MPG unknown asteroids dataset (red points), Swope (green) and INT (blue). 
          The three objects having the faintest $H$ are among the best NEO candidates. } 
  \label{fig8}
\end{figure*}

\begin{figure*}
  \centerline{
    \mbox{\includegraphics[width=7cm]{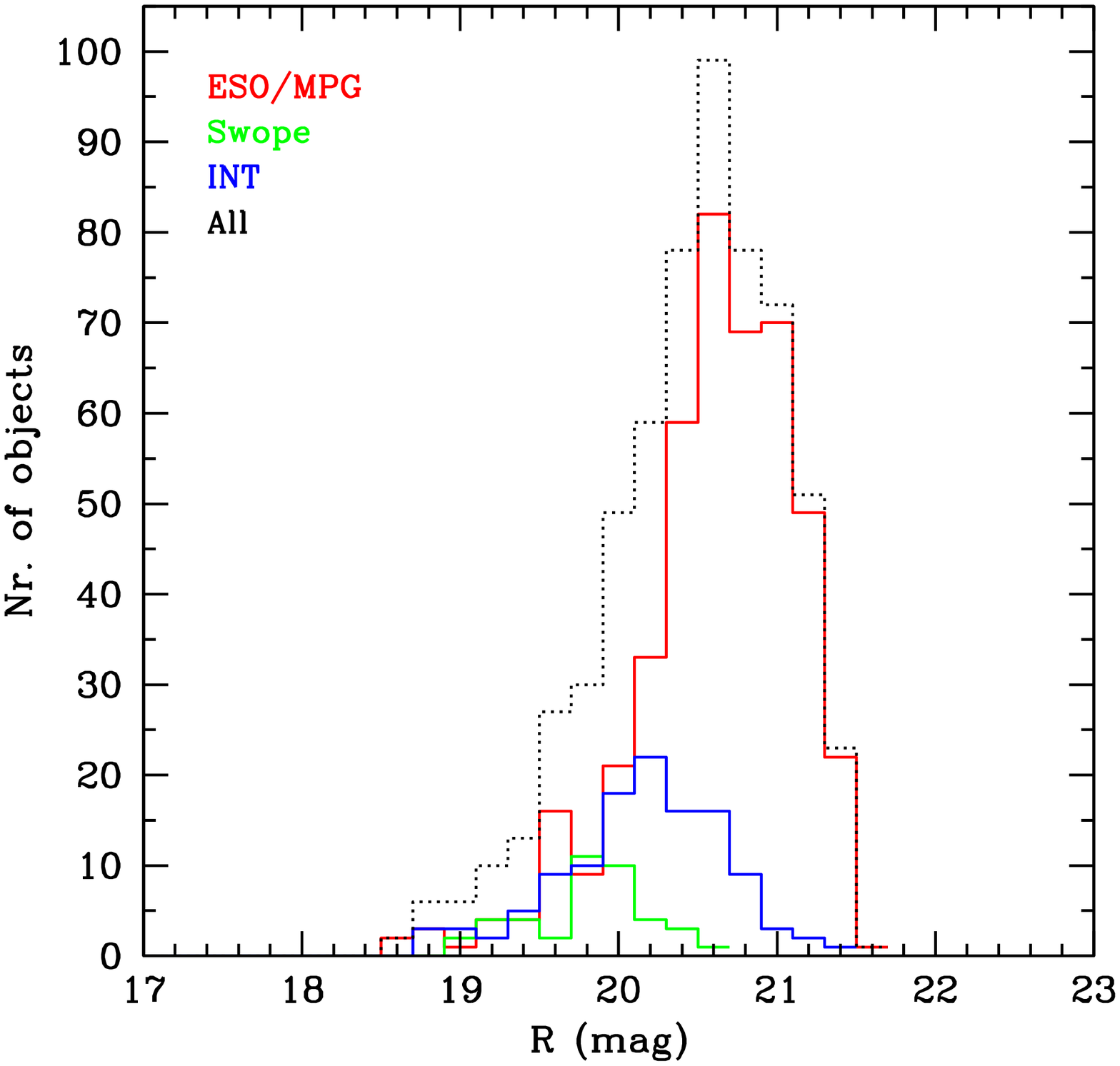}}
    \mbox{\includegraphics[width=7cm]{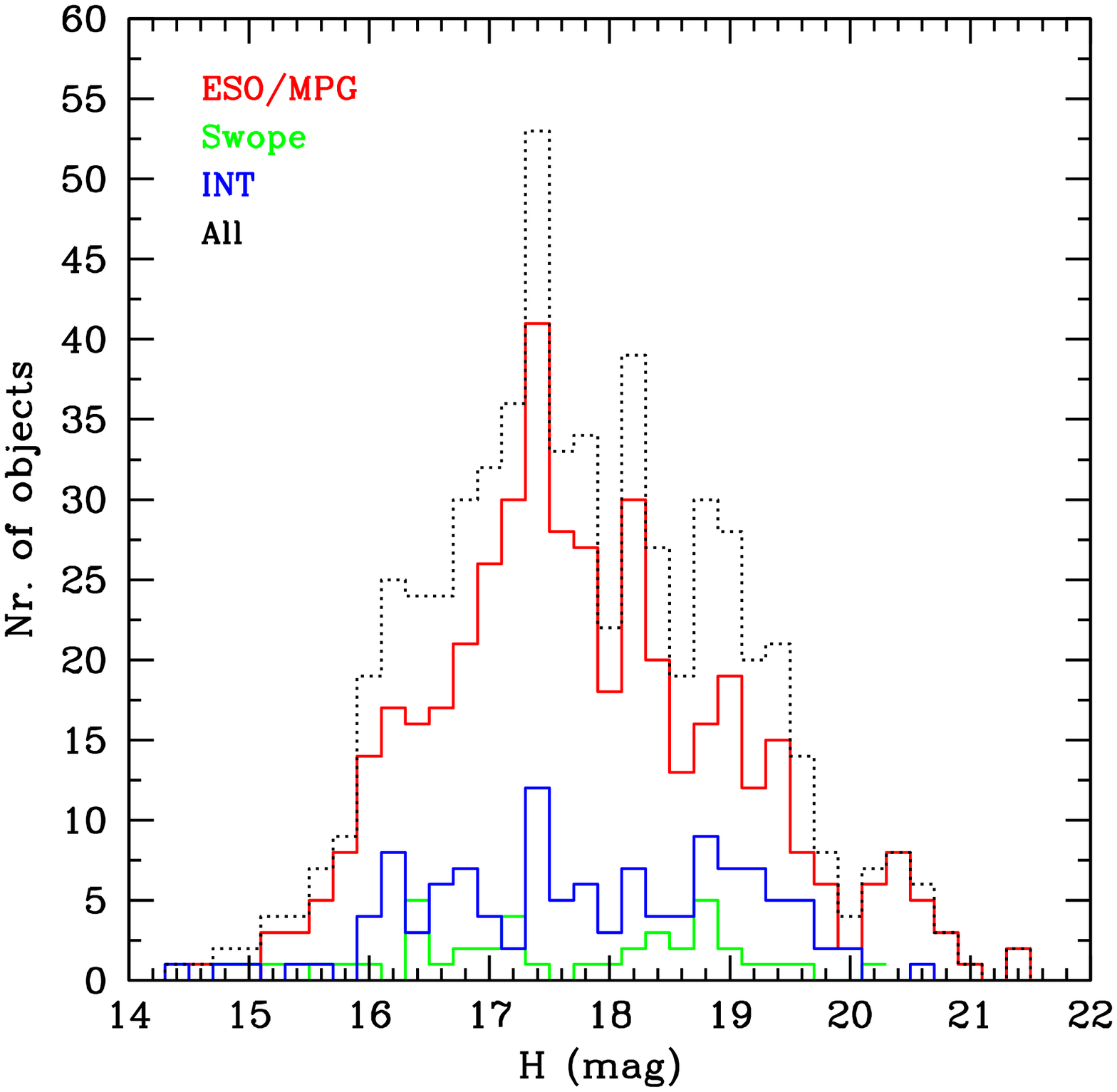}}
  }
  \caption{Histograms showing number of unknown objects as function of observed apparent $R$ magnitude 
          (left) and calculated absolute magnitude $H$ (right) for the ESO/MPG dataset (red), 
          Swope (green), INT (blue) and the total number of objects (black dots). } 
  \label{fig9}
\end{figure*}

The right panel of Figure~\ref{fig9} plots the histogram showing the calculated absolute 
magnitude $H$ for the unknown objects observed at the three facilities.  Five
objects fall outside the $H$ range of the plot, namely the brightest object
VB037 identified as the jovian Trojan 2001 TB234 ($H=13.3$) and the faintest
four objects VBTO016 ($H=22.7$), VBTU222 ($H=24.7$), VBVI213 ($H=25.3$) and
VTU021 ($H=26.4$).  The last four are visible as clear individual points in
Figure~\ref{fig8} (right) and the last three are among our best NEO
candidates (Table~\ref{table4}). The $H$
histograms are more evenly distributed, showing 2--3 maxima (possibly not all real) for each 
facility and an overall maximum at $H\sim17.4$ mag. This limit can be regarded as the limiting 
$H$ giving completeness for a 2m class facility for the entire main belt (including the outer 
region). According to \citet{yos07}, this limit corresponds to S-class asteroids about 1km in 
diameter, thus virtually all S-type MBAs larger than this limit should be accessible to a 2m 
telescope (including ESO/MPG and INT) in good weather conditions. 

As we saw in Figure~\ref{fig8} (left),
$R\sim21.6$ mag represents the limiting apparent magnitude for the ESO/MPG. 
Most MBAs have absolute magnitudes between $15 < H < 21$ mag, consistent with sizes between 
170m and 6km, assuming albedos between 0.05 and 0.25 \citep{nas11}. 
The four best NEO candidates have the following $H$ and sizes, assuming 
the same limits for their albedos:
VTU021:  $H=26.4$ mag, $13$--$31$ m;
VBVI213: $H=25.3$ mag, $22$--$55$ m;
VBTU222: $H=24.7$ mag, $30$--$70$ m;
VBTU197: $H=18.9$ mag, $440$--$980$ m. 

\begin{figure*}
  \centerline{
    \mbox{\includegraphics[width=7cm]{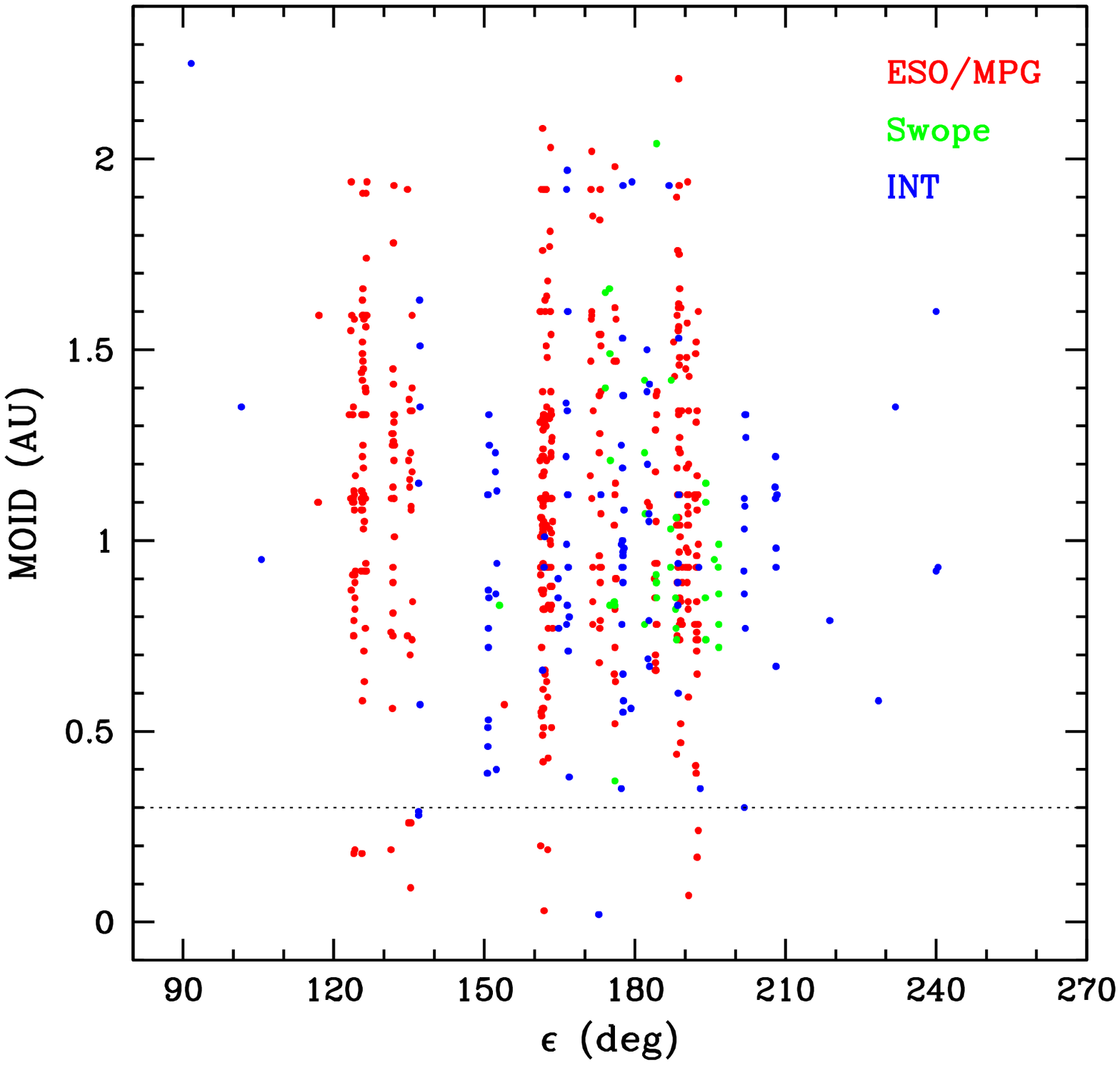}}
    \mbox{\includegraphics[width=7cm]{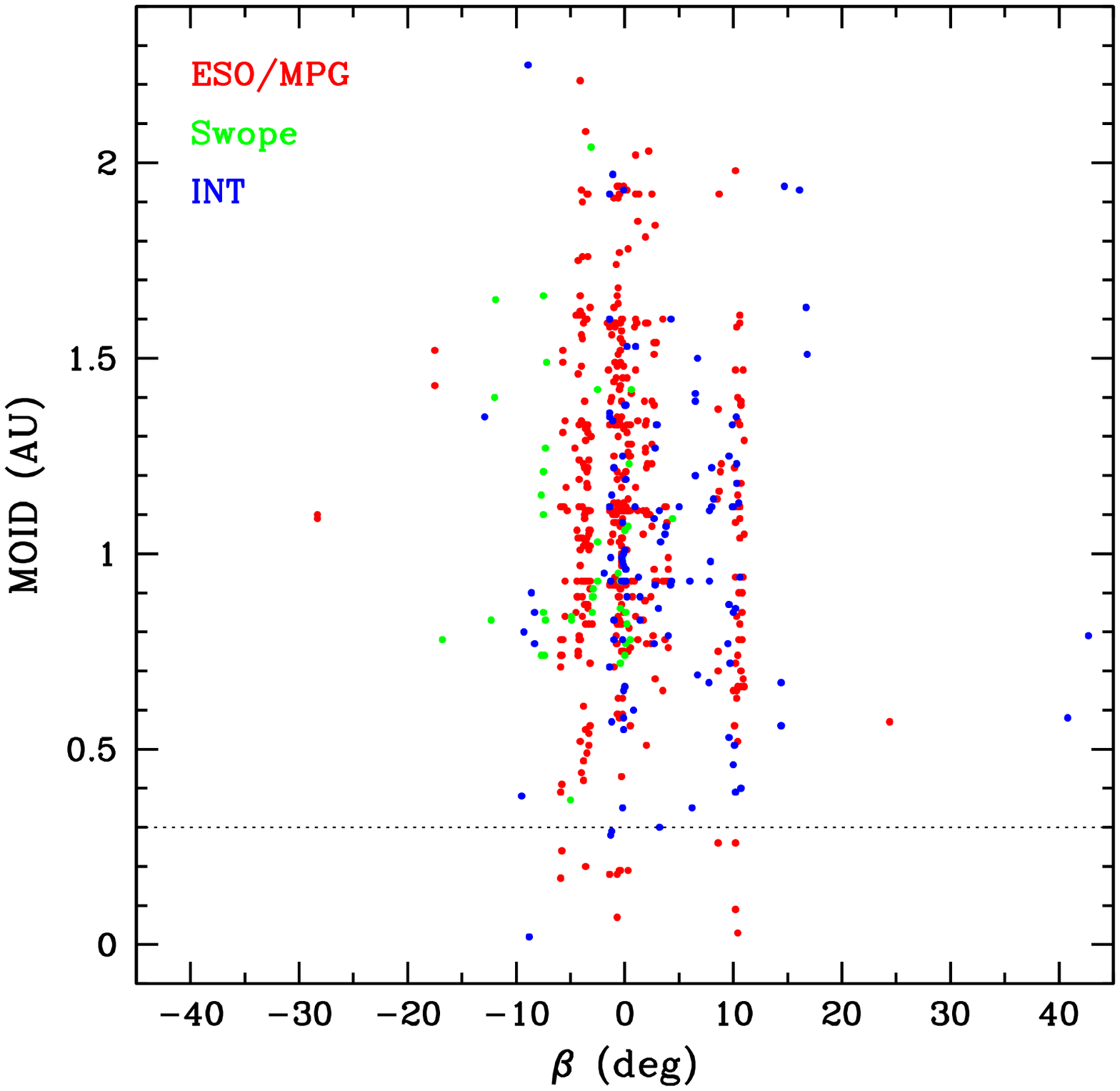}}
  }
  \caption{Minimal Orbital Intersection Distance (MOID) versus Solar elongation $\epsilon$ (left panel), 
	   and versus ecliptic latitude $\beta$ (right panel) for the unknown objects observed at ESO/MPG 
           (red), Swope (green) and INT (blue). The dotted line at $MOID<0.3$ marks the NEO region under which 
            all our NEO candidates appear. } 
  \label{fig10}
\end{figure*}

In the left panel of Figure~\ref{fig10} we plot the calculated MOID versus the elongation 
$\epsilon$ for the unknown objects observed at ESO/MPG (red), Swope (green) and INT (blue). 
With a dotted line we mark the MOID$=0.3$ limit for NEAs below which all our NEO candidates appear. 
Based on this plot, there is no apparent favorable elongation to discover NEOs. 

In the right panel of Figure~\ref{fig10} we plot the calculated MOID versus the observed 
ecliptic latitude $\beta$ for the unknown objects observed at ESO/MPG (red), Swope 
(green) and INT (blue). Most unknown objects and NEO candidates were observed at low 
latitudes, under $10^{\circ}$. Most NEO candidates are observed at low latitudes and 
there is no particular favorable detection latitude with respect of the MBAs. 

\subsection{Survey statistics for 2m and 1m facilities}
\label{stats}

Based on the statistics available from our ESO/MPG, Swope and INT surveys, we can evaluate 
the unknown MBA and NEA population observable at low latitudes ($\mid \beta \mid < 10^{\circ}$) 
by 2m and 1m surveys. We include these results in Table~\ref{table2}. 

\subsubsection{Unknown MBA density}

Using data from our ESO/MPG survey (the best performing 2m facility) we observed in the 
$10^{\circ}$ latitude range 347 known objects and 467 unknown objects scanning 13 square degrees.
This gives an average of 27 known and 36 unknown MBAs per square degree visible to 
limiting magnitude $R\sim21.5$ mag in 2 min exposure time using the ESO/MPG. We include 
these findings at the end of Table~\ref{table2}. These numbers give for ESO/MPG a MBA ratio 
known:unknown = 0.7. We compare below our findings with other authors. 

Counting data from our Swope survey we observed 35 unknown objects and 65 known objects 
within $10^{\circ}$ latitude range from the ecliptic scanning about 5 square degrees of sky.
This gives an average of 11 known objects and 7 unknown objects per square degree visible to 
limiting magnitude $R\sim20.4$ mag in 2 min exposure time with a 1m facility. The total number 
agrees with earlier results from the Spacewatch 0.9m which detected for the whole survey 16 
asteroids per square degree. 

\citet{boa04} conducted a 3 night pilot search and follow-up program to detect NEAs 
using the ESO/MPG with WFI in $3\times3$ binning mode (i.e., same facility and setup as us).
During the last two nights, the authors scanned in good weather conditions a total of 24 
square degrees, counting an average of 10 known and 12 unknown asteroids
(mostly MBAs) per square degree.
This gives a ratio known:unknown = 0.8 which is consistent with our findings. 
Nevertheless, both their numbers are about three times less than our findings. Their survey 
strategy was a bit different than ours, namely they observed at small solar elongation during 
the first and last part of the night and observing near opposition during the middle part. 
Also, for identification they used mostly automated software, although some data were
reduced with Astrometrica. It is well known that the eye and brain are better than computer 
software in detection of moving objects by a factor of 3/2 based on experience of Spacewatch II
\citep{boa04} or by $1-1.5$ mag according to other authors \citep{yos03}. Both these factors 
could explain the lower density of asteroids (mostly MBAs) found by \citet{boa04} compared with our 
statistics. 

\citet{wie07} searched 50 fields (50 square degrees) from the CFHT Legacy Survey (CFHTLS 3.6m) 
observed in $r^\prime$ close to opposition and within a $2^{\circ}$ latitude range from the ecliptic.
The moving objects were detected automatically by using the Sextractor software with a threshold 
$3\sigma$. The authors found an average of 70 asteroids per square degree up to 
$r^\prime \sim 21.5$ mag \citep{wie11}, which agrees well with our findings. 

Using the Subaru 8.3m telescope equipped with the large field SuprimeCam with 7 sec exposures 
\citet{yos03} surveyed 3 square degrees near opposition and the ecliptic (SMBAS I survey) and 
found 92 asteroids per square degree to limiting magnitude $R=21.5$ mag. Using the same facility
to image 4 square degrees using 2 second exposures (SMBAS II survey), \citet{yos07} found an 
average of 75 objects per square degree to the same limit. In these surveys, moving objects 
were detected by human inspection. Our findings using the ESO/MPG are very close to their 
densities, taking into account our lower S/N due to Subaru's larger aperture and their 
pointing at lower latitudes. 

Using one single 8.4m mirror of the Large Binocular Telescope (LBT), \citet{rya09} studied 
the asteroid distribution in the ecliptic, finding up to $V=22.3$ mag (close to $R\sim21.5$ mag 
our limit) a density of 85 asteroids per square degree. Asteroid detection was performed visually 
using a three-color method. Their density found is very close to ours, counting our total number 
of objects (63 objects per square degree). 

\subsubsection{Unkown NEA density}

Counting the NEO candidates from Table~\ref{table4}, ESO/MPG produced 8 NEO candidates 
and 3 best NEO candidates scanning a field of 13 square degrees. This gives between 0.2 
and 0.6 NEO candidates per square degree observable with this facility. The value 0.6
is an upper limit because we could not confirm our 
objects which were observed only in one night. 

Scanning 40 WFC fields (13 square degree) within $15^\circ$ latitude in good weather at 
ESO/MPG, \citet{boa04} discovered 3 NEA candidates (including 2 confirmed NEAs), which gives 
a density of 0.2 NEA candidates per square degree, matching our findings counting only
the best candidates. Comparing their results with those from the 1.8m Spaceguard II survey, 
the authors conclude that on average one NEA per 10 square degrees could be discovered with 
ESO/MPG and the WFI. This is consistent with our findings if we count only one object, 
namely our best NEA candidate, VBVI213. Counting all our NEO candidates, our result is 
6 times more optimistic than that of \citet{boa04}. 

According to Table~\ref{table4}, INT produced 8 unknown objects and only one best NEA 
candidate scanning a field of 10 square degrees. This gives between 0.1 and 0.8 NEO candidates 
per square degree observable with INT (mostly in bad conditions). These densities are similar 
with those found by ESO/MPG and consistent with any other 2m survey. 

\section{Conclusions}
\label{conclusions}

We have analysed our observations taken with the ESO/MPG 2.2m in La Silla,
the Swope 1m in Las Campanas and the INT 2.5m in La Palma.  The total sky
surveyed during 11 nights was about 29 square degrees, which allowed us to
study statistics of MBAs and NEAs observable nowadays by other 1--2m
facilities. Our main conclusions are:

\begin{itemize}

\item
These telescopes are successful at following up faint objects soon after
discovery, preventing their loss, recovering NEAs at their second or later 
opposition and eliminating NEA candidates and Virtual Impactors. 

\item
The majority of our unknown objects are consistent with MBAs, based on two
evaluation methods. Up to 16 unknown objects could represent NEO candidates
from which 4 represent our best NEO candidates according to three evaluation
methods.

\item
The O-C residuals for known MBAs and program NEAs amount to
$0.15^{\prime\prime}$ for the ESO/MPG, $0.39^{\prime\prime}$ and
$0.18^{\prime\prime}$ for Swope and $0.42^{\prime\prime}$ and
$0.66^{\prime\prime}$ for the INT, whose prime focus field is the most
distorted (especially the three non-central CCDs) and needs to be corrected
in order to improve the astrometry.

\item
The UCAC-2 catalog is better than USNO-B1 which shows an offset af
$0.1^{\prime\prime}$ to the North, consistent with previous findings of other
authors.

\item
Published orbits (specifically $a$, $e$ and $i$) of known asteroids are very
similar to our calculated orbits using the FIND\_ORB software based on our
observed very small arcs. 

\item
Based on statistics derived from our data, we could assess the observability of the unknown MBA 
and NEA populations using 1m and 2m class surveys. Employing a 1m facility one can observe today 
fewer unknown objects than known MBAs and virtually no new NEO. Using a 2m facility, a slightly 
larger number of unknown than known MBAs could be detected
(up to about $a=3.2$ AU), consistent with objects having sizes between 170m and 6km (taking into
account the limits of the main belt and the albedo range). Between 0.1 and 0.8 new NEO candidates 
per square degree could be discovered using a 2m telescope. 

\item
A basic model assuming circular and coplanar orbits of the asteroids and Earth could be used 
in order to check any large all sky survey for potential NEO candidates. Employing the proper 
motion and Solar elongations, this model does not depend on calculated quantities such as 
orbital elements possibly subject to errors. Compared with other tools such as the MPC's NEO
Rating and the calculated preliminary orbits, this model seems very accurate at small elongations
($\pm 30^{\circ}$ from opposition) but, based on the residuals in our data, smaller elongations
(around $120-140^{\circ}$) need further study.

\end{itemize}

\section{Acknowledgements}

This work was based on observations made with the ESO/MPG telescope at La Silla Observatory 
under programme ID 080.C-2003(A), the Swope telescope at Las Campanas Observatory (CNTAC 2008), 
both granted under Chilean time, 
and the INT telescope in La Palma under Director's Discretionary Time of Spain's Instituto 
de Astrofisica de Canarias (CAT DDT 2010). 
OV acknowledges ESO, LCO and IA/UCN for supporting the runs in Chile for himself and the students, 
and also to the ING, IAC and the IAA for supporting the run in La Palma for the students. 
OV and JL gratefully acknowledge support from the spanish ``Ministerio de Ciencia e 
Innovaci\'on'' (MICINN) project AYA2008-06202-C03-02. 
This research has made intensive use of the Astrometrica software developed by Herbert 
Raab, very simple to install and use by students and amateur astronomers. 
We also used the image viewer SAOImage DS9, developed by Smithsonian Astrophysical 
Observatory and also IRAF, distributed by the National Optical Astronomy Observatories, 
operated by the Association of Universities for Research in Astronomy, Inc. under cooperative 
agreement with the National Science Foundation. 
Special thanks are due to Bill Gray for providing FIND\_ORB and installation assistance. 
OV also acknowledges to Paul Wiegert for feedback necessary to compare MBA data observable 
with CFHT and the ESO/MPG telescopes. 
Thanks are due to Fumi Yoshida and Tsuko Nakamura for sharing their SMBAS data and interest. 
We also acknowledge to Jure Skvarc for his satellite identification software and to Lilian
Dominguez for providing some references about space debris.  
Thanks are also due to Alain Maury who helped us to count NEA discoveries made in Europe. 
Acknowledgements are due to the referee whose constructive suggestions helped us to improve the paper. 

\bibliographystyle{elsarticle}

% ___________________________________________
%
% FIGURES  FIGURES  FIGURES  FIGURES  FIGURES 
% ___________________________________________

\newpage

%________________________________________________

% MAIN TABLES (INCL IN PAPER)
%________________________________________________

\newpage
\renewcommand{\arraystretch}{0.8}
\begin{table}[!t]
\begin{center}
\caption{The observing log for NEA observations at INT. We list the name of the asteroid, 
its classification at time of observation,
date of observation, expected apparent magnitude $V$, exposure time (seconds), 
number of observed positions, apparent motion $\mu$ ($^{\prime\prime}$/min), ephemeris uncertainty 
(arcsec) and observed orbital arc since discovery (d-days, m-months, y-years). Objects marked 
with * represent special cases discussed in the paper. } 
\label{table1}
{\small
\begin{tabular}{lcrrrrrrr}
\noalign{\smallskip}
\noalign{\smallskip}
\hline
\hline
\noalign{\smallskip}
\smallskip
Asteroid  &  Class  &  Date     &    $V$  &  Exp   &  Nr   &           $\mu$         &           $3\sigma$  &  Obs  \\
          &         &    (UT)   &  (mag)  &  (s)   &  pos  & ($^{\prime\prime}/min$) & ($^{\prime\prime}$)  &  arc  \\
\hline
\noalign{\smallskip}
\noalign{\smallskip}
2009 CW1   &  NEA  &  2009 Feb 12  &  19.9  &  20  &  10  &   3  &   40 &   9d  \\
2009 CB2*  &   ?   &  2009 Feb 12  &  20.1  &  60  &   5  &   3  & 1000 &   6d  \\
2009 CA2   &  NEA  &  2009 Feb 12  &  20.0  &  30  &   8  &   1  &   30 &   9d  \\
2003 SJ84* &  NEA  &  2009 Feb 12  &  21.4  & 120  &   5  &   2  &   81 &   6y  \\
\noalign{\smallskip}
\noalign{\smallskip}
2009 FF    &  PHA  &  2009 Apr 15  &  20.8  &  60  &   5  &   2  &   38 &   1m  \\
2009 DZ    &  PHA  &  2009 Apr 15  &  21.1  & 120  &   6  &   1  &    3 &   1m  \\
\hline
\hline
\end{tabular}
}
\end{center}
\end{table}

\newpage
\begin{table}[!t]
\begin{center}
\caption{Summary of the ESO/MPG, Swope and the INT runs. } 
\label{table2}
{\small
\begin{tabular}{lrrrr}
\noalign{\smallskip}
\noalign{\smallskip}
\hline
\hline
\noalign{\smallskip}
\smallskip
Observations &  ESO/MPG  &  Swope &  INT  &  Total  \\
\hline
\noalign{\smallskip}
\noalign{\smallskip}
Program NEAs                               &    15   &    50  &    35   &  100  \\
Nr of positions                            &   156   &   506  &   169   &  831  \\
O--C standard deviation ($^{\prime\prime}$)&  0.15   &  0.39  &  0.42   &  ---  \\
\noalign{\smallskip}
\noalign{\smallskip}
Known MBAs                                 &   347   &    68  &   143   &  558  \\
Nr of positions                            & 2,976   &   680  &   733   & 4,389 \\
O--C standard deviation ($^{\prime\prime}$)&  0.15   &  0.18  &  0.66   &  ---  \\
\noalign{\smallskip}
\noalign{\smallskip}
Unknown objects                            &   467   &    41  &   120   &  628  \\
\hline
\hline
\end{tabular}
}
\end{center}
\end{table}

\newpage
\begin{table}[!t]
\begin{center}
\caption{Minor Planet Circulars (MPC) and Minor Planet Electronic Circulars (MPEC) publishing our NEA observations} 
\label{table3}
{\footnotesize
\begin{tabular}{lll}
\noalign{\smallskip}
\noalign{\smallskip}
\hline
\hline
\noalign{\smallskip}
Telescope &  MP(E)C  &  Reference \\
\hline
\noalign{\smallskip}
\noalign{\smallskip}
Swope    &  MPC 64484, 5   &  Benecchi et al (2008a)  \\
Swope    &  MPC 64096, 9   &  Benecchi et al (2008b) \\
Swope    &  MPC 66688, 10  &  Benecchi et al (2009a) \\
Swope    &  MPC 64753, 3   &  Benecchi et al (2009b) \\
Swope    &  MPC 63365, 10  &  Kern et al (2008) \\
Swope    &  MPC 66190, 11  &  Kern et al (2009) \\
Swope    &  MPEC 2008-U48  &  Pozo et al (2008) \\
Swope    &  MPEC 2008-U46  &  Scotti et al (2008a) \\
Swope    &  MPEC 2008-U45  &  Scotti et al (2008b) \\
Swope    &  MPEC 2008-K66  &  Tubbiolo et al (2008) \\
Swope    &  MPC 63125, 6   &  Vaduvescu \& Tudorica (2008) \\
Swope    &  MPEC 2008-K63  &  Young et al (2009) \\
\noalign{\smallskip}
\noalign{\smallskip}
ESO/MPG  &  MPC 63369, 9   &  Cavadore et al (2009) \\
ESO/MPG  &  MPC 63591, 10  &  Elst et al (2008a) \\
ESO/MPG  &  MPC 63129, 5   &  Elst et al (2008b) \\
ESO/MPG  &  MPC 62871, 8   &  Elst et al (2008c) \\
ESO/MPG  &  MPC 62573, 3   &  Elst et al (2008d) \\
ESO/MPG  &  MPC 66195, 4   &  Elst et al (2009a) \\
ESO/MPG  &  MPC 65331, 2   &  Elst et al (2009b) \\
ESO/MPG  &  MPC 65045, 7   &  Elst et al (2009c) \\
ESO/MPG  &  MPC 62258, 1   &  Sheppard et al (2008) \\
ESO/MPG  &  MPC 62262, 5   &  Tholen et al (2008) \\ 
ESO/MPG  &  MPC 65636, 2   &  Tholen et al (2009) \\
\noalign{\smallskip}
\noalign{\smallskip}
INT      &   MPC 66196, 2  &  Holman et al (2009) \\
INT      &   MPEC 2010-E39 &  Holvorcem et al (2010) \\
INT      &   MPC 66457, 4  &  Fitzsimmons et al. (2009a) \\
INT      &   MPC 65927, 9  &  Fitzsimmons et al. (2009b) \\
INT      &   MPC 65332, 1  &  Fitzsimmons et al. (2009c) \\
\noalign{\smallskip}
\noalign{\smallskip}

\hline
\hline
\end{tabular}
}
\end{center}
\end{table}

\newpage
\begin{table}[!t]
\begin{center}
\caption{Near Earth Object (NEO) candidates -- fast unknown objects observed at ESO/MPG (first group) and the INT (second group). 
We list in bold 4 objects which qualify as the best NEO candidates, having small MOIDs, $100\%$ NEO Rating score and 
fitting best the $\epsilon-\mu$ model. The two one-nighter objects marked with * are not NEAs according to their later identifications.}
\label{table4}
{\footnotesize
\begin{tabular}{lrrrrrrrrrrr}
\noalign{\smallskip}
\noalign{\smallskip}
\hline
\hline
\noalign{\smallskip}
Acronym &             $\mu$       &  $\epsilon$ &  $R$   &  MOID  &  $a$  &  $e$  &      $i$   &  Nr  &          $\sigma$    &   MPC    &  Model  \\
        & ($^{\prime\prime}$/min) &  ($^\circ$) &        &  (AU)  & (AU)  &       & ($^\circ$) & pos  & ($^{\prime\prime}$)  &   score  &         \\ 
\hline
\noalign{\smallskip}
\noalign{\smallskip}
VB011         &        0.55  &  125  &      20.8  &       0.19  &  2.28  &  0.49  &   6.7  &   8  &   0.13  &         14  &     close  \\
VBTU203       &        0.65  &  132  &      20.8  &       0.75  &  1.84  &  0.05  &  23.5  &  16  &   0.26  &         10  &       fit  \\
{\bf VBTU197} &        0.68  &  135  &      21.4  &  {\bf 0.09} &  3.02  &  0.68  &   8.2  &   7  &   0.38  &  {\bf 100}  & {\bf best} \\
VBVA002       &        0.60  &  154  &      19.8  &       0.57  &  2.50  &  0.39  &  13.8  &   3  &   0.08  &         22  &       fit  \\
{\bf VBTU222} &  {\bf 2.12}  &  162  &      20.4  &  {\bf 0.03} &  1.01  &  0.08  &   2.2  &   5  &   0.16  &  {\bf 100}  & {\bf best} \\
VBTU226       &        0.89  &  162  &      21.0  &       0.56  &  1.94  &  0.20  &  25.7  &   5  &   0.52  &         21  &       fit  \\
\hline
\hline
\end{tabular}
}
\end{center}
\end{table}

\end{document}

% ______________________________

% APPENDIX TABLES
% ______________________________

% ESO/MPG DATA 
%__________

\newpage
\appendix
\section{Unknown objects observed during out runs at ESO/MPG, Swope and INT. } 

\scriptsize
\begin{center}
\begin{longtable}{llrrrrrrrrrrrr}
%Here is the caption, the stuff in [] is the table of contents entry,
%the stuff in {} is the title that will appear on the first page of the
%table.
\caption{The asteroids officialy discovered at ESO/MPG (according to MPC DISCSTATUS Jan 2011). In the first line we give orbital elements 
calculated with FIND\_ORB and observational data from our run, while in the second line we include MPC orbital data derived from all available observations. } 
\label{tableA1}
\\
%This is the header for the first page of the table...
\hline \hline \\
   \multicolumn{1}{c}{Acronym} &
   \multicolumn{1}{c}{Designation} &
   \multicolumn{1}{c}{$a$} &
   \multicolumn{1}{c}{$e$} &
   \multicolumn{1}{c}{$i$} &
   \multicolumn{1}{c}{MOID} &
   \multicolumn{1}{c}{$H$} &
   \multicolumn{1}{c}{pos} &
   \multicolumn{1}{c}{arc} &
   \multicolumn{1}{c}{$rms$} &
   \multicolumn{1}{c}{$R$} &
   \multicolumn{1}{c}{$\beta$} &
   \multicolumn{1}{c}{$\epsilon$} &
   \multicolumn{1}{c}{$\mu$} \\

%   \multicolumn{1}{c}{} &
%   \multicolumn{1}{c}{} &
%   \multicolumn{1}{c}{(AU)} &
%   \multicolumn{1}{c}{} &
%   \multicolumn{1}{c}{($^\circ$)} &
%   \multicolumn{1}{c}{(AU)} &
%   \multicolumn{1}{c}{mag} &
%   \multicolumn{1}{c}{} &
%   \multicolumn{1}{c}{} &
%   \multicolumn{1}{c}{($^{\prime\prime}$)} &
%   \multicolumn{1}{c}{(mag)} &
%   \multicolumn{1}{c}{$^\circ$} &
%   \multicolumn{1}{c}{$^\circ$} &
%   \multicolumn{1}{c}{$^{\prime\prime}$/min} \\

\\ \hline \hline \\
\endfirsthead

%This is the header for the remaining page(s) of the table...
\multicolumn{14}{c}{{\tablename} \thetable{} (continued) -- Asteroids officially discovered at ESO/MPG. } \\
\\  \hline \hline \\
   \multicolumn{1}{c}{Acronym} &
   \multicolumn{1}{c}{Designation} &
   \multicolumn{1}{c}{$a$} &
   \multicolumn{1}{c}{$e$} &
   \multicolumn{1}{c}{$i$} &
   \multicolumn{1}{c}{MOID} &
   \multicolumn{1}{c}{$H$} &
   \multicolumn{1}{c}{pos} &
   \multicolumn{1}{c}{arc} &
   \multicolumn{1}{c}{$\sigma$} &
   \multicolumn{1}{c}{$R$} &
   \multicolumn{1}{c}{$\beta$} &
   \multicolumn{1}{c}{$\epsilon$} &
   \multicolumn{1}{c}{$\mu$} \\
\\ \hline \hline \\
\endhead

VBSO004 & 2008 EB98 & 2.94 & 0.12 & 7.7 & 1.58 & 16.8 & 18 & 2d & 0.07 & 20.0 & 10 & 176 & 0.54 \\
 &  & 2.90 & 0.16 & 7.1 &  & 17.2 & 25 & 10d &  &  &  &  &  \\
VBSO008 & 2008 EF155 & 2.73 & 0.05 & 11.4 & 1.61 & 17.3 & 16 & 2d & 0.14 & 20.5 & 11 & 176 & 0.60 \\
 &  &  &  &  &  &  &  &  &  &  &  &  &  \\
VBSO036 & 2008 EG162 & 2.86 & 0.21 & 1.9 & 1.28 & 18.5 & 7 & 1d & 0.24 & 20.7 & 2 & 173 & 0.54 \\
 &  & 2.79 & 0.18 & 2.3 &  & 18.2 & 15 & 23d &  &  &  &  &  \\
VBSO052 & 2008 EV168 & 2.49 & 0.06 & 3.3 & 1.33 & 17.5 & 8 & 1d & 0.10 & 20.1 & 2 & 163 & 0.57 \\
 &  & 2.41 & 0.16 & 2.7 &  & 18.2 & 16 & 24d &  &  &  &  &  \\
\end{longtable}
\end{center}

% _________

\newpage
\scriptsize
\begin{center}
\begin{longtable}{llrrrrrrrrrrrr}
%Here is the caption, the stuff in [] is the table of contents entry,
%the stuff in {} is the title that will appear on the first page of the
%table.
\caption{Later identification of unknown asteroids observed at ESO/MPG. First line represent our data and second line reffers to updated MPC data. } 
\label{tableA2} 
\\
%This is the header for the first page of the table...
\hline \hline \\
   \multicolumn{1}{c}{Acronym} &
   \multicolumn{1}{c}{Designation} &
   \multicolumn{1}{c}{$a$} &
   \multicolumn{1}{c}{$e$} &
   \multicolumn{1}{c}{$i$} &
   \multicolumn{1}{c}{MOID} &
   \multicolumn{1}{c}{$H$} &
   \multicolumn{1}{c}{pos} &
   \multicolumn{1}{c}{arc} &
   \multicolumn{1}{c}{$rms$} &
   \multicolumn{1}{c}{$R$} &
   \multicolumn{1}{c}{$\beta$} &
   \multicolumn{1}{c}{$\epsilon$} &
   \multicolumn{1}{c}{$\mu$} \\
\\ \hline \hline \\
\endfirsthead

%This is the header for the remaining page(s) of the table...
\multicolumn{14}{c}{{\tablename} \thetable{} (continued) -- Later identification of unknown asteroids observed at ESO/MPG. } \\
\\  \hline \hline \\
   \multicolumn{1}{c}{Acronym} &
   \multicolumn{1}{c}{Designation} &
   \multicolumn{1}{c}{$a$} &
   \multicolumn{1}{c}{$e$} &
   \multicolumn{1}{c}{$i$} &
   \multicolumn{1}{c}{MOID} &
   \multicolumn{1}{c}{$H$} &
   \multicolumn{1}{c}{pos} &
   \multicolumn{1}{c}{arc} &
   \multicolumn{1}{c}{$rms$} &
   \multicolumn{1}{c}{$R$} &
   \multicolumn{1}{c}{$\beta$} &
   \multicolumn{1}{c}{$\epsilon$} &
   \multicolumn{1}{c}{$\mu$} \\
\\ \hline \hline \\
\endhead

VB001 & 2008 FG67 & 2.52 & 0.16 & 6.0 & 1.11 & 17.3 & 8 & 1d & 0.27 & 20.4 & 0 & 124 & 0.23 \\
 &  & 2.68 & 0.22 & 7.2 &  & 17.5 & 28 & 48d &  &  &  &  &  \\
VB014 & 2004 RZ319 & 2.22 & 0.23 & 7.4 & 0.75 & 16.7 & 8 & 1d & 0.07 & 20.0 & 0 & 124 & 0.14 \\
 &  & 3.16 & 0.02 & 14.2 &  & 15.3 & 77 & 11y &  &  &  &  &  \\
VB018 & 2008 FZ110 & 2.63 & 0.12 & 4.5 & 1.33 & 16.6 & 8 & 1d & 0.08 & 20.1 & 0 & 124 & 0.13 \\
 &  & 2.52 & 0.10 & 4.6 &  & 17.2 & 46 & 9y &  &  &  &  &  \\
VB024 & 2009 TG3 & 2.21 & 0.05 & 5.1 & 1.10 & 16.8 & 8 & 1d & 0.07 & 20.4 & -1 & 124 & 0.11 \\
 &  & 2.62 & 0.21 & 7.2 &  & 16.5 & 47 & 7y &  &  &  &  &  \\
\end{longtable}
\end{center}

% _________

\newpage
\scriptsize
\begin{center}
\begin{longtable}{lrrrrrrrrrrrr}
%Here is the caption, the stuff in [] is the table of contents entry,
%the stuff in {} is the title that will appear on the first page of the
%table.
\caption{Outstanding unknown objects observed at ESO/MPG. } 
\label{tableA3} 
\\
%This is the header for the first page of the table...
\hline \hline \\
   \multicolumn{1}{c}{Acronym} &
   \multicolumn{1}{c}{$a$} &
   \multicolumn{1}{c}{$e$} &
   \multicolumn{1}{c}{$i$} &
   \multicolumn{1}{c}{MOID} &
   \multicolumn{1}{c}{$H$} &
   \multicolumn{1}{c}{pos} &
   \multicolumn{1}{c}{arc} &
   \multicolumn{1}{c}{$rms$} &
   \multicolumn{1}{c}{$R$} &
   \multicolumn{1}{c}{$\beta$} &
   \multicolumn{1}{c}{$\epsilon$} &
   \multicolumn{1}{c}{$\mu$} \\
\\ \hline \hline \\
\endfirsthead

%This is the header for the remaining page(s) of the table...
\multicolumn{13}{c}{{\tablename} \thetable{} (continued) -- Outstanding unknown objects observed at ESO/MPG. } \\
\\  \hline \hline \\
   \multicolumn{1}{c}{Acronym} &
   \multicolumn{1}{c}{$a$} &
   \multicolumn{1}{c}{$e$} &
   \multicolumn{1}{c}{$i$} &
   \multicolumn{1}{c}{MOID} &
   \multicolumn{1}{c}{$H$} &
   \multicolumn{1}{c}{pos} &
   \multicolumn{1}{c}{arc} &
   \multicolumn{1}{c}{$rms$} &
   \multicolumn{1}{c}{$R$} &
   \multicolumn{1}{c}{$\beta$} &
   \multicolumn{1}{c}{$\epsilon$} &
   \multicolumn{1}{c}{$\mu$} \\
\\ \hline \hline \\
\endhead

VB002 & 3.12 & 0.06 & 8.4 & 1.94 & 15.9 & 8 & 1d & 0.16 & 21.1 & 0 & 123 & 0.09 \\
VB003 & 2.38 & 0.11 & 3.0 & 1.11 & 17.7 & 8 & 1d & 0.08 & 20.8 & 0 & 123 & 0.19 \\
VB004 & 2.62 & 0.19 & 14.8 & 1.11 & 17.6 & 8 & 1d & 0.16 & 20.6 & 0 & 124 & 0.38 \\
VB005 & 3.03 & 0.29 & 2.1 & 1.13 & 17.1 & 7 & 1d & 0.09 & 20.6 & 0 & 124 & 0.24 \\
VB006 & 2.22 & 0.05 & 6.5 & 1.12 & 17.1 & 8 & 1d & 0.10 & 20.6 & 0 & 124 & 0.11 \\
VB007 & 2.29 & 0.16 & 1.1 & 0.92 & 18.2 & 8 & 1d & 0.14 & 20.8 & 0 & 124 & 0.26 \\
VB008 & 2.38 & 0.23 & 2.3 & 0.85 & 16.3 & 7 & 1d & 0.16 & 20.9 & 0 & 124 & 0.10 \\
VB010 & 2.85 & 0.23 & 2.3 & 1.17 & 17.0 & 8 & 1d & 0.11 & 21.0 & 0 & 124 & 0.11 \\
\end{longtable}
\end{center}

% _________

% SWOPE DATA 
% _________

\newpage
\scriptsize
\begin{center}
\begin{longtable}{llrrrrrrrrrrrr}
%Here is the caption, the stuff in [] is the table of contents entry,
%the stuff in {} is the title that will appear on the first page of the
%table.
\caption{Later identification of unknown asteroids observed with Swope telescope. First line represent our data and second line reffers to updated MPC data. } 
\label{tableA4} 
\\
%This is the header for the first page of the table...
\hline \hline \\
   \multicolumn{1}{c}{Acronym} &
   \multicolumn{1}{c}{Designation} &
   \multicolumn{1}{c}{$a$} &
   \multicolumn{1}{c}{$e$} &
   \multicolumn{1}{c}{$i$} &
   \multicolumn{1}{c}{MOID} &
   \multicolumn{1}{c}{$H$} &
   \multicolumn{1}{c}{pos} &
   \multicolumn{1}{c}{arc} &
   \multicolumn{1}{c}{$rms$} &
   \multicolumn{1}{c}{$R$} &
   \multicolumn{1}{c}{$\beta$} &
   \multicolumn{1}{c}{$\epsilon$} &
   \multicolumn{1}{c}{$\mu$} \\
\\ \hline \hline \\
\endfirsthead

PBS007 & 2008 UC335 & 2.44 & 0.27 & 3.2 & 0.78 & 19.3 & 5 & 1d & 0.27 & 19.9 & 1 & 182 & 0.53 \\
 &  & 2.14 & 0.25 & 1.9 &  & 20.2 & 11 & 18d &  &  &  &  &  \\
PBS009 & 2008 UE35 & 2.27 & 0.09 & 5.2 & 1.07 & 17.1 & 8 & 1d & 0.22 & 19.8 & 0 & 182 & 0.67 \\
 &  & 2.21 & 0.1 & 3.1 &  & 18.3 & 30 & 36d &  &  &  &  &  \\
PBS015 & 2008 SM192 & 2.38 & 0.23 & 3.7 & 0.83 & 18.6 & 10 & 1d & 0.22 & 19.5 & -7 & 175 & 0.59 \\
 &  & 2.8 & 0.14 & 5.4 &  & 16.9 & 20 & 30d &  &  &  &  &  \\
PBT010 & 2008 TZ101 & 2.48 & 0.17 & 1.7 & 1.06 & 18.0 & 7 & 1d & 0.12 & 19.8 & 0 & 188 & 0.59 \\
 &  & 2.42 & 0.19 & 1.5 &  & 18.5 & 38 & 48d &  &  &  &  &  \\
\end{longtable}
\end{center}

%_______

\newpage
\scriptsize
\begin{center}
\begin{longtable}{lrrrrrrrrrrrr}
%Here is the caption, the stuff in [] is the table of contents entry,
%the stuff in {} is the title that will appear on the first page of the
%table.
\caption{Outstanding unknown objects observed with Swope telescope. } 
\label{tableA5} 
\\
%This is the header for the first page of the table...
\hline \hline \\
   \multicolumn{1}{c}{Acronym} &
   \multicolumn{1}{c}{$a$} &
   \multicolumn{1}{c}{$e$} &
   \multicolumn{1}{c}{$i$} &
   \multicolumn{1}{c}{MOID} &
   \multicolumn{1}{c}{$H$} &
   \multicolumn{1}{c}{pos} &
   \multicolumn{1}{c}{arc} &
   \multicolumn{1}{c}{$rms$} &
   \multicolumn{1}{c}{$R$} &
   \multicolumn{1}{c}{$\beta$} &
   \multicolumn{1}{c}{$\epsilon$} &
   \multicolumn{1}{c}{$\mu$} \\
\\ \hline \hline \\
\endfirsthead

PBS001 & 2.20 & 0.21 & 4.1 & 0.74 & 19.5 & 6 & 1d & 0.23 & 20.5 & -7 & 194 & 0.52 \\
PBS002 & 2.44 & 0.24 & 3.6 & 0.85 & 19.3 & 6 & 1d & 0.14 & 20.6 & -7 & 194 & 0.46 \\
PBS003 & 2.40 & 0.13 & 4.2 & 1.10 & 18.3 & 6 & 1d & 0.18 & 20.4 & -7 & 194 & 0.53 \\
PBS004 & 2.50 & 0.16 & 9.4 & 1.15 & 18.0 & 6 & 1d & 0.28 & 20.4 & -8 & 194 & 0.59 \\
PBS005 & 2.20 & 0.21 & 4.3 & 0.74 & 18.8 & 6 & 1d & 0.07 & 19.8 & -8 & 194 & 0.52 \\
PBS006 & 2.49 & 0.03 & 5.4 & 1.42 & 17.2 & 8 & 1d & 0.25 & 19.9 & 1 & 182 & 0.62 \\
PBS008 & 2.72 & 0.18 & 6.0 & 1.23 & 17.2 & 8 & 1d & 0.15 & 19.3 & 0 & 182 & 0.60 \\
PBS010 & 2.24 & 0.19 & 6.1 & 0.83 & 18.8 & 8 & 1d & 0.17 & 20.3 & -12 & 153 & 0.46 \\
\end{longtable}
\end{center}

% _________

% INT DATA 
% _________

\newpage
\scriptsize
\begin{center}
\begin{longtable}{llrrrrrrrrrrrr}
%Here is the caption, the stuff in [] is the table of contents entry,
%the stuff in {} is the title that will appear on the first page of the
%table.
\caption{Later identification of unknown asteroids observed at the INT. First line represent our data and second line reffers to updated MPC data. } 
\label{tableA6} 
\\
%This is the header for the first page of the table...
\hline \hline \\
   \multicolumn{1}{c}{Acronym} &
   \multicolumn{1}{c}{Designation} &
   \multicolumn{1}{c}{$a$} &
   \multicolumn{1}{c}{$e$} &
   \multicolumn{1}{c}{$i$} &
   \multicolumn{1}{c}{MOID} &
   \multicolumn{1}{c}{$H$} &
   \multicolumn{1}{c}{pos} &
   \multicolumn{1}{c}{arc} &
   \multicolumn{1}{c}{$rms$} &
   \multicolumn{1}{c}{$R$} &
   \multicolumn{1}{c}{$\beta$} &
   \multicolumn{1}{c}{$\epsilon$} &
   \multicolumn{1}{c}{$\mu$} \\
\\ \hline \hline \\
\endfirsthead

%This is the header for the remaining page(s) of the table...
\multicolumn{14}{c}{{\tablename} \thetable{} (continued) -- Later identification of unknown asteroids observed at the INT. } \\
\\  \hline \hline \\
   \multicolumn{1}{c}{Acronym} &
   \multicolumn{1}{c}{Designation} &
   \multicolumn{1}{c}{$a$} &
   \multicolumn{1}{c}{$e$} &
   \multicolumn{1}{c}{$i$} &
   \multicolumn{1}{c}{MOID} &
   \multicolumn{1}{c}{$H$} &
   \multicolumn{1}{c}{pos} &
   \multicolumn{1}{c}{arc} &
   \multicolumn{1}{c}{$rms$} &
   \multicolumn{1}{c}{$R$} &
   \multicolumn{1}{c}{$\beta$} &
   \multicolumn{1}{c}{$\epsilon$} &
   \multicolumn{1}{c}{$\mu$} \\
\\ \hline \hline \\
\endhead

VDT070 & 247902 & 2.30 & 0.08 & 6.4 & 1.12 & 17.8 & 4 & 1d & 0.19 & 20.3 & 10 & 151 & 0.37 \\
 &  & 2.86 & 0.20 & 9.0 &  & 16.3 & 51 & 10y &  &  &  &  &  \\
VDT072 & 2005 YW147 & 2.38 & 0.11 & 6.3 & 1.12 & 17.3 & 5 & 1d & 0.41 & 19.7 & 10 & 151 & 0.39 \\
 &  & 2.42 & 0.15 & 5.8 &  & 17.6 & 33 & 6y &  &  &  &  &  \\
VDT073 & 2008 SZ203 & 2.43 & 0.13 & 2.7 & 1.12 & 17.8 & 2 & 1d & 0.00 & 19.7 & 5 & 173 & 0.61 \\
 &  & 2.52 & 0.06 & 3.4 &  & 17.2 & 30 & 3y &  &  &  &  &  \\
VIT003 & 2005 TO93 & 2.77 & 0.06 & 6.5 & 1.60 & 16.2 & 4 & 1d & 0.42 & 20.4 & 4 & 240 & 0.17 \\
 &  & 2.60 & 0.10 & 8.3 &  & 16.9 & 53 & 14y &  &  &  &  &  \\
\end{longtable}
\end{center}

%_______

\newpage
\scriptsize
\begin{center}
\begin{longtable}{lrrrrrrrrrrrr}
%Here is the caption, the stuff in [] is the table of contents entry,
%the stuff in {} is the title that will appear on the first page of the
%table.
\caption{Outstanding unknown objects observed at the INT. } 
\label{tableA7} 
\\
%This is the header for the first page of the table...
\hline \hline \\
   \multicolumn{1}{c}{Acronym} &
   \multicolumn{1}{c}{$a$} &
   \multicolumn{1}{c}{$e$} &
   \multicolumn{1}{c}{$i$} &
   \multicolumn{1}{c}{MOID} &
   \multicolumn{1}{c}{$H$} &
   \multicolumn{1}{c}{pos} &
   \multicolumn{1}{c}{arc} &
   \multicolumn{1}{c}{$rms$} &
   \multicolumn{1}{c}{$R$} &
   \multicolumn{1}{c}{$\beta$} &
   \multicolumn{1}{c}{$\epsilon$} &
   \multicolumn{1}{c}{$\mu$} \\
\\ \hline \hline \\
\endfirsthead

%This is the header for the remaining page(s) of the table...
\multicolumn{13}{c}{{\tablename} \thetable{} (continued) -- Outstanding unknown objects observed at the INT. } \\
\\  \hline \hline \\
   \multicolumn{1}{c}{Acronym} &
   \multicolumn{1}{c}{$a$} &
   \multicolumn{1}{c}{$e$} &
   \multicolumn{1}{c}{$i$} &
   \multicolumn{1}{c}{MOID} &
   \multicolumn{1}{c}{$H$} &
   \multicolumn{1}{c}{pos} &
   \multicolumn{1}{c}{arc} &
   \multicolumn{1}{c}{$rms$} &
   \multicolumn{1}{c}{$R$} &
   \multicolumn{1}{c}{$\beta$} &
   \multicolumn{1}{c}{$\epsilon$} &
   \multicolumn{1}{c}{$\mu$} \\
\\ \hline \hline \\
\endhead

VBA001 & 2.26 & 0.13 & 5.6 & 0.98 & 18.2 & 5 & 1d & 0.23 & 20.2 & 8 & 208 & 0.45 \\
VBA002 & 2.72 & 0.23 & 11.0 & 1.11 & 18.4 & 5 & 1d & 0.22 & 20.8 & 8 & 208 & 0.51 \\
VBA003 & 1.80 & 0.02 & 23.5 & 0.79 & 19.0 & 5 & 1d & 0.08 & 21.1 & 43 & 219 & 0.45 \\
VBA004 & 2.66 & 0.30 & 11.7 & 0.90 & 17.4 & 3 & 1d & 0.53 & 20.2 & -9 & 165 & 0.67 \\
VBA005 & 2.65 & 0.31 & 4.7 & 0.85 & 17.8 & 5 & 1d & 0.85 & 20.2 & -8 & 165 & 0.68 \\
VBA006 & 2.27 & 0.22 & 4.6 & 0.77 & 18.7 & 5 & 1d & 0.26 & 19.7 & -8 & 165 & 0.47 \\
VDT055 & 2.61 & 0.00 & 2.3 & 1.60 & 17.4 & 5 & 1d & 0.24 & 20.6 & -1 & 167 & 0.57 \\
VDT056 & 2.38 & 0.11 & 6.0 & 1.12 & 18.7 & 3 & 1d & 0.17 & 20.8 & -1 & 167 & 0.62 \\
\end{longtable}
\end{center}